\UseRawInputEncoding
\documentclass[12pt,draftclsnofoot,onecolumn]{IEEEtran}

\usepackage{amssymb}
\usepackage{amsthm,amsmath,amssymb}
\usepackage{mathrsfs}
\usepackage{amsfonts}
\usepackage{indentfirst}
\usepackage{graphicx}
\usepackage{subfigure}
\usepackage{float}
\usepackage{subeqnarray}
\usepackage{cases}
\usepackage{bm}
\usepackage{textcomp}
\usepackage{cite}
\usepackage{color}
\usepackage{diagbox}
\usepackage[table,xcdraw]{xcolor}
\usepackage{pifont}

\ifCLASSINFOpdf

\else

\fi

\makeatletter
\newif\if@restonecol
\makeatother

\usepackage[linesnumbered,ruled,vlined]{algorithm2e}
\usepackage{algpseudocode}

\newtheorem{theorem}{Theorem}
\newtheorem{lemma}{Lemma}
\newtheorem{assumption}{Assumption}
\newtheorem{definition}{Definition}

\begin{document}

\title{Energy-Efficient Wireless Federated Learning via Doubly Adaptive Quantization}

\author{Xuefeng Han, Wen Chen, Jun Li, Ming Ding, Qingqing Wu, \\ Kang Wei, Xiumei Deng, Zhen Mei
\thanks{Xuefeng Han, Wen Chen and Qingqing Wu are with the Broadband Access Network Laboratory, Shanghai Jiao Tong University, Minhang 200240, China
(e-mail: \{hansjell-watson; wenchen; qingqingwu\}@sjtu.edu.cn).
Jun Li, Xiumei Deng and Zhen Mei are with School of Electrical and Optical Engineering, Nanjing University of Science and Technology, Nanjing 210094, China Ministry of Education (e-mail: \{jun.li; xiumeideng; meizhen\}@njust.edu.cn).
Ming Ding is with Data61, CSIRO, Sydney, NSW 2015, Australia (e-mail:
ming.ding@data61.csiro.au).
Kang Wei is with the Department of Computing, Hong Kong Polytechnic
University, Hong Kong, 999077, China (e-mail: kangwei@polyu.edu.hk).}}


\maketitle

\begin{abstract}
Federated learning (FL) has been recognized as a viable distributed learning paradigm for training a machine learning model across distributed clients without uploading raw data. However, FL in wireless networks still faces two major challenges, i.e., large communication overhead and high energy consumption, which are exacerbated by client heterogeneity in dataset sizes and wireless channels. While model quantization is effective for energy reduction, existing works ignore adapting quantization to heterogeneous clients and FL convergence. To address these challenges, this paper develops an energy optimization problem of jointly designing \underline{q}uantization levels, scheduling \underline{c}lients, allocating \underline{c}hannels, and controlling computation \underline{f}requencies (QCCF) in wireless FL. Specifically, we derive an upper bound identifying the influence of client scheduling and quantization errors on FL convergence. Under the long-term convergence constraints and wireless constraints, the problem is established and transformed into an instantaneous problem with Lyapunov optimization. Solving Karush-Kuhn-Tucker conditions, our closed-form solution indicates that the doubly adaptive quantization level rises with the training process and correlates negatively with dataset sizes. Experiment results validate our theoretical results, showing that QCCF consumes less energy with faster convergence compared with state-of-the-art baselines.
\end{abstract}


\IEEEpeerreviewmaketitle

\section{Introduction}
Integrated Artificial Intelligence (AI) and Communication have been identified as the usage scenarios of the sixth generation mobile networks. AI algorithms are widely utilized in numerous mobile devices and play a crucial role in various aspects of life, such as AI healthcare and autonomous vehicles, thanks to leaping computation ability and increasing data storage capacity. However, traditional training of AI models requires large amounts of data, and the issue of data privacy has recently attracted widespread public attention. As a distributed training paradigm, federated learning (FL) has been proposed by Google in \cite{google}. Unlike centralized learning, a server shares a model, and distributed clients collaboratively train the shared model without any raw data exchange in FL, which preserves data privacy of clients. Also, FL can be integrated with existing techniques, for instance, differential privacy \cite{TIFS} and blockchain \cite{blockchain_FL} to enhance security.
\par
Due to the pervasive and scalable connectivity of wireless networks, FL is integrated with wireless networks, in which the server and clients exchange models through downlink and uplink. Different from the adequately resourced server, energy consumption in wireless FL is considerable for clients. Furthermore, a stringent latency requirement for each communication round compels clients to increase the computation frequency and uplink power to accomplish computation and communication in time, leading to more energy consumption. Among clients, moreover, heterogeneity in energy arrivals \cite{online_policy}, mobilities \cite{mobility}, and dataset sizes \cite{different_epoch} is also common in wireless FL. Different dataset sizes, as the focus of this article, make the allocation of identical resources to each client impact client participating and waste energy in vain.
\par
To address the aforementioned dilemma, several lightweight methods for the model have been proposed to ease the burden of computation and communication. For instance, model pruning \cite{model_pruning} sets unnecessary parameters of the model to zero, and model distillation \cite{layer_distillation} allows clients to train simple student models under the guidance of a complex teacher model. Also, model partition \cite{model_partition} divides the model into two parts, with one part in the client and the other in the server. These methods effectively reduce the model size, leading to energy savings and latency reduction. In addition to the above methods, quantization is also validated for reducing energy consumption. It quantizes parameters of models or updates with fewer bits, without additional knowledge such as masks of pruning or the teacher model, nor other steps like loss propagation between divided models. However, the attendant quantization error has an impact on FL convergence. In this context, to adapt to both the convergent training process and heterogeneous clients with different dataset sizes, our doubly adaptive quantization for wireless FL is proposed.
\subsection{Related Works}
The goal of energy optimization has spurred various efforts that schedule clients and allocate resources such as central processing unit (CPU) frequencies, transmitting power, and channels. For instance, \cite{incentive} controlled the number of participating clients in subsequent communication rounds based on the accuracy of the current aggregated global model. \cite{schedule_clients} scheduled clients based on their energy efficient ratios. Whereas, \cite{incentive, schedule_clients} only focused on FL and overlooked concrete wireless scenes. Based on this point, \cite{NOMA} focused on the non-orthogonal multiple access system, and \cite{outage} considered the probabilistic outage of channels. Unfortunately, high transmitting power or CPU frequencies decided by resource allocation methods in \cite{NOMA, outage} still consume enormous energy with unchanged communication overhead and cannot alleviate the heterogeneity.
\par
Now we turn to the model quantization. There are many quantization methods integrated with FL in \cite{binary_neural_network, knob_choose, block_floating_point, encode_quantization, pruning_quantization, model_distillation}. With the lowest quantization level, binary neural networks were applied in the FL framework and the corresponding convergence was derived in \cite{binary_neural_network}. And with a higher quantization level, \cite{knob_choose} proposed a quantization knob choosing method for the fixed-point quantization. Consisting of mantissa and exponent, the block floating point quantization method is utilized in \cite{block_floating_point}. \cite{encode_quantization} mapped a local model to a bit sequence according to an encoding matrix and reconstructed the model in the server. In addition to the aforementioned quantization methods, other lightweight methods can also be synergistically combined. For example, in \cite{pruning_quantization}, model pruning and quantization were applied before and after local updates, respectively. Moreover, \cite{model_distillation} employed a combination of model distillation and soft quantization.
\par
The above works broadened the scope of applying model quantization in FL. Notwithstanding, the majority of follow-up works adopted a fixed quantization level rather than a varying quantization level, which has the full potential of communication overhead reduction and energy savings due to the varying tolerance of the model for quantization errors in the whole process. \cite{quantization_rise} proposed a rising quantization level with the training process according to its derived optimal quantization level. On the contrary,  in \cite{maximun_element}, the quantization level was determined by the L-$\infty$ norm of local updates and was decreased with the training process. In essence, whether the quantization level rises or descends, it is evident that an appropriate quantization level depends on the current model and varies with the training process. With uploading local updates, a criterion about quantization levels was established to schedule clients in \cite{Fedpaq}. However, \cite{quantization_rise, Fedpaq} both ignored wireless constraints and \cite{maximun_element} even set 70-second latency for each communication round, which deviates significantly from real latency over wireless networks. After all, it is wireless constraints that aggravate the impact of the heterogeneity. Without a specific quantization design for clients, the timeout or high energy consumption of clients with large datasets cannot completely be removed by simply adapting the quantization level to the training process.
\par
To this end, it becomes imperative for the quantization level to adapt to clients. In \cite{quantization_criterion}, different criteria of quantization levels were designed for different clients. In addition, \cite{client_time_adaptive} adopted time-client-adaptive quantization levels, anchored in the global loss function value and incrementally adjusted according to local dataset sizes. \cite{both_computation_communication} applied an adaptive model quantization method in both computation and communication. Regrettably, despite the insightful analyses of FL performance in \cite{quantization_criterion, client_time_adaptive, both_computation_communication}, wireless constraints and energy consumption were largely overlooked. Hence, \cite{Lyapunov_quantization} defined the excess risk of the channel and the mean-square deviation to design an adaptive quantization level. Similarly, \cite{correlated_Gaussian} transformed local gradients into correlated Gaussian random variables which were adaptively quantized, and controlled the communication rate accordingly. Notwithstanding, \cite{Lyapunov_quantization, correlated_Gaussian} focused on different communication rates among clients and gave no adaption to different sizes due to the same-size dataset setting.
\par
To sum up, none of the previous works provides insights into the relationships of the quantization level with both the training process and different dataset sizes, i.e., doubly adaptive quantization. Additionally, an integration of doubly adaptive quantization and methods about resource allocation and client scheduling over wireless networks is notably absent. Hence, there is no energy-efficient wireless FL framework under this condition yet.
\subsection{Contributions}
This paper proposes an energy-efficient FL framework that addresses the heterogeneity problem caused by different dataset sizes over wireless networks. The key focus of the FL framework lies in the double adaption of quantization to both the training process and different dataset sizes owned by clients. Our main contributions are summarized as follows.
\begin{itemize}
  \item We derive an upper bound of gradients with the dataset sizes, the quantization levels, and the result of client scheduling. The upper bound related to FL convergence is divided into a data property part and a quantization error part. Under the long-term constraints of the two parts, we formulate an energy optimization problem by jointly designing \underline{q}uantization levels, scheduling \underline{c}lients, allocating \underline{c}hannels, and controlling computation \underline{f}requencies (QCCF).
  \item With Lyapunov optimization, the above long-term optimization problem is transformed into an equivalent instantaneous optimization problem of QCCF in each communication round. We then decompose the equivalent problem into two subproblems: the combinatorial subproblem is solved by a genetic algorithm, while the continuous subproblem, proven to be convex, possesses a closed-form solution according to Karush-Kuhn-Tucker (KKT) conditions. The closed-form solution suggests that the quantization level should gradually rise with the training process and be negatively correlated to the dataset size.
  \item Extensive experimental results demonstrate that our QCCF algorithm converges faster and sharply reduces the energy consumption by 48.21\% and 35.42\% compared with the principle algorithm in \cite{client_time_adaptive} and the same-size algorithm in \cite{Lyapunov_quantization}. As the divergence of dataset sizes increases, the advantage of our QCCF algorithm becomes more pronounced. Furthermore, the results of quantization levels validate the above derived relationships of the quantization level with both the training process and dataset sizes.
\end{itemize} \par
The rest of this article is organized as follows. First, our FL framework and the quantization method are described in Section II. Based on the FL framework, the corresponding convergence analysis is provided in Section III. Section IV gives the physical system of wireless FL and formulates the optimization problem. Then, the problem is solved in Section V and numerical results of experiments are described in Section VI. Finally, Section VII concludes this paper.
\section{Federated Learning with Model Quantization}
\subsection{Federated Learning Process}
\begin{figure*}[t]
  \centering
  \includegraphics[width=0.9\textwidth]{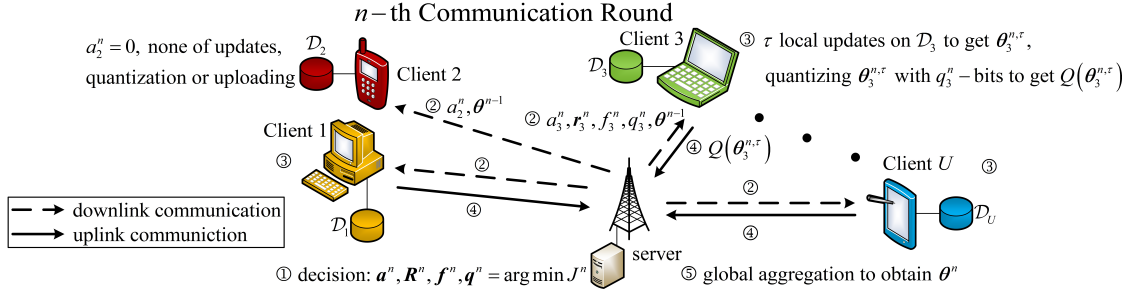}
  \caption{There are 5 steps in each communication round of our FL framework.}
  \label{figure: FL Architecture}
\end{figure*}

Consider a cellular network concluding $U$ clients and one server to accomplish an FL task in $N$ communication rounds. $ \mathcal{U} = \{1,2,\cdots, U\}$ denotes the set of all clients. As Fig. \ref{figure: FL Architecture} shows, there are five steps in the $n$-th communication round as follows. \par
\subsubsection{Decision}
The server generates decision variables $\bm a^n, \bm R^n, \bm f^n, \bm q^n$ and schedules clients to participate in the $n$-th communication round. $\bm a^n = [a_1^n, \cdots, a_U^n]$ is a participant indicator vector, where $a_i^n \in \{0, 1\}$ denotes the participant state of client $i$ with $a_i^n = 1$ indicating client $i$ participating in $n$-th communication round, and vice versa. Hence, participating clients form $\mathcal U^n = \{i | a_i^n = 1\}$. Other variables $\bm R^n = [\bm r_1^n, \cdots, \bm r_U^n], \bm f^n = [f_1^n, \cdots, f_U^n],$ and $\bm q^n = [q_1^n, \cdots, q_U^n]$ will be introduced later and the detailed decision process will be provided in Section V.
\subsubsection{Broadcasting}
In the downlink communication, the server broadcasts decision variables and the global model $\bm \theta^{n-1}$ obtained in the last communication round to all clients. In the first communication round, the initial model $\bm \theta^0$ is broadcast.  \par
\subsubsection{Local updating and Quantization}
There are neither updates nor quantization for any client out of $\mathcal U^n$. As for client $i \in \mathcal U^n$, local updates are executed on its dataset $\mathcal D_i$ with an assigned CPU frequency $f_i^n$. Specifically, client $i$ starts with $\bm \theta^{n, 0}_i = \bm \theta^{n-1},$
and $\bm \theta_i^{n,m}$ is updated by the mini-batch stochastic gradient of the local loss function $F_i(\cdot)$ in $m$-th update as
\begin{equation}
\bm\theta^{n, m+1}_i = \bm\theta^{n, m}_i - \eta \nabla F_i(\bm\theta^{n, m}_i, \xi_i^{n, m}), \enspace m = 0, 1, \cdots, \tau-1,
\label{equation: local update}
\end{equation}
where $\eta$ is the learning rate, $\xi_i^{n,m}$ is the sampled mini-batch, and $\tau$ is the number of local updates for all clients. Since there are $\tau^{\rm e}$ epochs within $\tau$ local updates, $\tau$ is an integral multiple of $\tau^{\rm e}$. After $\tau$ local updates, $\bm \theta_i^{n, \tau}$ is obtained and then quantized by the quantization function $Q(\cdot)$ with $q_i^n$-bits to get $Q(\bm \theta_i^{n, \tau})$. The specific quantization method will be introduced in Section II-B. \par
\subsubsection{Uploading}
In the uplink communication, client $i \in \mathcal U^n$ uploads $Q(\bm \theta_i^{n, \tau})$ with assigned channel $\bm r^n_i$ to the server.
\subsubsection{Global aggregation}
Quantized local models are aggregated to obtain the global model as
\begin{equation}
\boldsymbol{\theta}^{n} = \sum_{i =1}^U w_i^n Q(\boldsymbol{\theta}^{n, \tau}_{i}) ,
\label{equation: global aggregation}
\end{equation}
where $w_i^n = \frac{a_i^nD_i}{D^n}$ is the aggregation weight. $D_i$ denotes the size of $\mathcal D_i$, and $D^n = \sum_{i =1}^U a_i^n D_i$. \par
After $N$ communication rounds, the final global model $\boldsymbol{\theta}^{N}$ is obtained and needs to perform well on current datasets. Namely, the loss function of $\bm \theta^N$ on all local datasets is expected to tend to its minimum. Denoting $w_i = \frac{D_i}{\sum^U_{j=1} D_j}$, the final global loss function is written by
\begin{equation}
\min_{\boldsymbol{\theta}^{N}} \left( F(\boldsymbol{\theta}^{N}) \triangleq \sum^U_{i=1}w_i F_i ({\boldsymbol{\theta}^N})\right).
\label{equation: initial aim}
\end{equation}
\par
(\ref{equation: local update}) and (\ref{equation: global aggregation}) tell that $\boldsymbol{\theta}^{n}$ is firmly related to the previous global model, the quantization method, and client scheduling. To minimize $F(\bm \theta^N)$ in (\ref{equation: initial aim}), it is vital to choose $(\bm q^n, \bm a^n)$ for $n=1, \cdots, N.$
\subsection{Quantization Method}
In FL without model quantization, it is normally assumed that the server can get accurate local models. Limited to communication resources, however, it is not feasible to transmit accurate local models. Therefore, the quantization method \cite{stochastic_quantization} is adopted to tackle this issue.
\par
Firstly, the range of $\bm\theta^{n, \tau}_i \in \mathbb R^Z$ with $Z$ dimensions is measured by $\theta ^{n, \max}_i = \max \{ |\theta_{i,z}^{n, \tau} | \big| z = 1,2, \cdots, Z\},$
where $\theta_{i,z}^{n, \tau}$ denotes the $z$-th dimension of $\bm \theta_i^{n, \tau}$. With the $q_i^n$-bit stochastic quantization method for each dimension, $\theta^{n, \max}_i$ is divided into $2^{q_i^n} - 1$ intervals whose knobs are $k_u = \frac{u \theta^{n, \max}_i}{2^{q_i^n} - 1}, u = 0, 1, \cdots, 2^{q_i^n} - 1.$ And we can simply find the $u$-th interval which contains the $z$-th dimension of the model, that is, $|\theta_{i,z}^{n, \tau}| \in [k_u, k_{u+1})$. Hence, the quantized result of $\theta_{i,z}^{n, \tau}$ is
\begin{equation}
Q(\theta_{i,z}^{n, \tau}) = \left\{ \begin{array}{ll}
{\rm sign}(\theta_{i,z}^{n, \tau}) \cdot k_u, & {\rm with\ probability\ } \frac{k_{u+1} - |\theta_{i,z}^{n, \tau}|}{k_{u+1} - k_u}, \\
{\rm sign}(\theta_{i,z}^{n, \tau}) \cdot k_{u+1}, & {\rm with\ probability\ } \frac{|\theta_{i,z}^{n, \tau}| - k_u}{k_{u+1} - k_u}.
\end{array} \right.
\label{equation: quantization result}
\end{equation}\par
Quantized results of all dimensions of the model stack into the quantized model $Q(\bm \theta^{n, \tau}_i) = [Q(\theta^{n, \tau}_{i, 1}), \cdots, Q(\theta^{n, \tau}_{i, Z}) ]^\top$, which is uploaded to the server. According to the above quantization method, it is noticed that transmitted data about $Q(\bm \theta^{n, \tau}_i)$ includes three parts: the range, signs, and indexes of quantization knobs. To sum up, the bit length of $Q(\bm \theta^{n, \tau}_i)$ is computed by
\begin{equation}
\ell_i^n = Z q^n_i + Z + 32,
\end{equation}
where 32 means that the range is expressed by a 32-bit floating point number and $Z$ represents all 1-bit signs of $Z$ dimensions. Although no accurate local model can be obtained, the quantized model, as a substitute, has unbiasedness and boundedness as follows.
\begin{lemma}
For the quantization function $Q(\cdot)$, the accurate local model is unbiasedly estimated by
$\mathbb E[Q(\bm \theta^{n, \tau}_i)] = \bm \theta^{n, \tau}_i,$
and the variance is bounded by
$\mathbb E\left[ \left\|Q(\bm \theta^{n, \tau}_i) - \bm \theta^{n, \tau}_i \right\|^2 \right] \leq \frac{Z (\theta_{i}^{n, \max})^2}{4(2^{q_i^n} - 1)^2}.$
\label{lemma: quantization property}
\end{lemma}
\begin{IEEEproof}
The above two equations and proofs can be seen in (7) of \cite{quantization_rise}.
\end{IEEEproof}
\textbf{Lemma \ref{lemma: quantization property}} reveals that the quantization result can also represent the accurate model within the margin of the quantization error, which decreases as the quantization level $q_i^n$ rises.
\section{Convergence Analysis\label{section: convergence}}
In this section, we derive the upper bound of accumulated gradients to reflect the influences of scheduling and quantization levels on the model convergence to guarantee FL performance.
\subsection{Assumptions and Definitions}
As the premise, there are \textbf{Assumptions \ref{assumption: gradient}-\ref{assumption: mini-batch}}, which indicate the boundedness, the smoothness, and the sample property of the local gradient, respectively. Meanwhile, we also define a virtual global model in \textbf{Definition \ref{defintion}} to act as a bridge between local models and the global model.
\begin{assumption}
The norm of a local gradient at each client has an upper bound, i.e., $\forall i \in \mathcal U$, $\| \nabla F_i(\bm \theta^{n, m}) \| \leq G_i^n,$ for $ n = 1, 2, \cdots, N,$ and $m = 0, 1, \cdots, \tau$.
\label{assumption: gradient}
\end{assumption}
\begin{assumption}
The local loss functions of all clients are differentiable, and are $L$-smooth, i.e., $\| \nabla F_i(\bm \theta) - \nabla F_i(\bm \phi) \| \leq L \| \bm \theta - \bm \phi \|,$ for each $ i \in \mathcal U$ and any two models $\bm \theta, \bm \phi \in \mathbb R^Z$.
\label{assumption: smooth}
\end{assumption}
\begin{assumption}
The gradient on a mini-batch is an unbiased estimator and its variance can be bounded, i.e., $\forall i \in \mathcal U$, there are $\mathbb E [\nabla F_i(\bm \theta_i^{n,m}, \xi_i^{n,m})] = \mathbb E [\nabla F_i(\bm \theta_i^{n,m})]$ and $ \mathbb E[\| \nabla F_i(\bm \theta_i^{n,m}, \xi_i^{n,m})-\nabla F_i(\bm \theta_i^{n,m}) \|^2] \leq (\sigma_i^n)^2 $ for $ n = 1, 2, \cdots, N,$ and $m = 0, 1, \cdots, \tau$.
\label{assumption: mini-batch}
\end{assumption}
\begin{definition}
The virtual global model comes from aggregating local models of participating clients, i.e., $\bm \psi^{n,m} \triangleq \sum_{i \in \mathcal U^n} w_i^n \bm \theta^{n,m}_i$. When $m=0$, we have $\bm \psi^{n,0} \triangleq \sum_{i \in \mathcal U^n}  w_i^n \bm \theta^{n,0}_i = \bm \theta^{n-1}.$
\label{defintion}
\end{definition}
\subsection{Results}
As a base, local models at $m$-th update are first considered in \textbf{Lemma \ref{lemma: differences}}.
\begin{lemma}
If $2\eta^2 m^2 L^2 < 1$, the sum of differences between local models and the initial model is bounded by
\begin{equation}
\sum^U_{i = 1} w_i^n \mathbb E \left[ \| \bm \theta_i^{n,m} - \bm \psi^{n,0} \|^2 \right] \leq \frac{\eta^2m \sum^U_{i = 1} w_i^n (\sigma_i^n)^2 + 2\eta^2m^2 \sum^U_{i=1} w_i^n (G_i^n)^2}{1 - 2 \eta^2 m^2 L^2}.
\label{equation: upper bound model}
\end{equation}
\label{lemma: differences}
\end{lemma}
\begin{IEEEproof}
Please refer to the detailed proof in Appendix \ref{proof1}.
\end{IEEEproof}
In \textbf{Lemma \ref{lemma: differences}}, we observe that the upper bound rises as $m$ rises, and it will vanish to 0 when $m=0$. These corollaries align with realities. As for the gradient, we have \textbf{Theorem \ref{theorem: update}}.
\begin{theorem}
If $\eta L < 1$, the gradient norm of the virtual global model at the $m$-th update in the $n$-th communication round has an upper bound as
\begin{equation}
\begin{aligned}
{\mathbb E}\left[ {\left\| {\nabla F\left( {\bm \psi ^{n,m} } \right)} \right\|^2 } \right] \leq & \frac{2}{\eta }{\mathbb E}\left[ {F\left( {\bm \psi ^{n,m} } \right) - F\left( {\bm \psi ^{n,m + 1} } \right)} \right] + 4\sum\limits_{i = 1}^U {(1 - a_i^n w_i) \left( {G_i^n } \right)^2 } \\
& + \eta L\sum\limits_{i = 1}^U {w_i^n \left( {\sigma_i^n } \right)^2 } + \frac{{2\eta ^2 mL^2 \sum\limits_{i = 1}^U {w_i^n \left( {\sigma _i^n } \right)^2 }  + 4\eta ^2 m^2 L^2 \sum\limits_{i = 1}^U {w_i^n \left( {G_i^n } \right)^2 } }}{{1 - 2\eta ^2 \tau ^2 L^2 }}.
\label{equation: upper bound update}
\end{aligned}
\end{equation}
\label{theorem: update}
\end{theorem}
\begin{IEEEproof}
Please refer to the detailed proof in Appendix \ref{proof2}.
\end{IEEEproof}
From \textbf{Theorem \ref{theorem: update}}, we note that the norm of the gradient is firmly related to the data property represented by $\sigma_i^n, G_i^n$ and client participation represented by $w_i^n$. However,  there is no term related to quantization error in \textbf{Theorem \ref{theorem: update}} at one update. Hence, another upper bound considering the quantization method across all communication rounds is given in \textbf{Theorem \ref{theorem: round}}.
\begin{theorem}
If $2\eta^2 \tau^2 L^2 < 1$, the sum of virtual global model gradient norms is bounded by
\begin{small}
\begin{align}
& \sum\limits_{n = 0}^{N-1} {\sum\limits_{m = 0}^{\tau-1} {{\mathbb E}\left[ {\left\| {\nabla F\left( {\bm \psi ^{n,m} } \right)} \right\|^2 } \right]} }   \leq \frac{2}
{\eta }{\mathbb E}\left[ {F\left( {\bm\theta ^0 } \right) - F\left( {\bm\theta ^N } \right)} \right] + \frac{L}
{2}\sum\limits_{n = 1}^{N} {\sum\limits_{i = 1}^U {w_i^n \frac{ Z ( \theta_{i}^{n, \max})^2 }{4(2^{q_i^n} - 1)^2} }}   + \eta L\tau\sum\limits_{n = 1}^{N} \sum^U_{i=1} w_i^n (\sigma_i^n)^2 \notag  \\
& \quad + \eta^2 L^2 \sum\limits_{n = 1}^{N} {\frac{{\left( {\tau^2 - \tau} \right) \sum\limits_{i = 1}^U {w_i^n \left( {\sigma _i^n } \right)^2 }  + \frac{2 (2\tau^3 - 3\tau^2 + \tau) }{3} \sum\limits_{i = 1}^U {w_i^n \left( {G_i^n } \right)^2 } }}{{1 - 2\eta ^2 \tau ^2 L^2 }}} + 4\tau \sum\limits_{n = 1}^{N} {\sum\limits_{i = 1}^U {(1-a_i^nw_i)\left( {G_i^n } \right)^2 } }. \label{equation: upper bound round}
\end{align}
\end{small}
\label{theorem: round}
\end{theorem}
\begin{IEEEproof}
Summing up (\ref{equation: upper bound update}) for $m = 0, 1, \cdots, \tau-1$ and reducing denominator $1 - 2\eta^2m^2L^2$ to $1 - 2\eta^2\tau^2L^2$, we have
\begin{equation}
\begin{aligned}
& \sum^{\tau - 1}_{m = 0}\mathbb E \left[\left\| \nabla F(\bm \psi^{n,m}) \right\|^2\right] \leq \frac{2}{\eta} \mathbb E[F(\bm \psi^{n, 0}) - F(\bm \psi^{n, \tau})] + 4 \tau \sum^U_{i=1} (1 - a_i^n w_i) (G_i^n)^2\\
& \quad  + \eta \tau L \sum^U_{i=1} w_i^n (\sigma_i^n)^2 + \eta^2 L^2\frac{(\tau^2 - \tau) \sum^U_{i=1} w_i^n (\sigma_i^n)^2 + \frac{2(2\tau^3 - 3\tau^2 + \tau)}{3} \sum^U_{i=1} w_i^n (G_i^n)^2}{1 - 2 \eta^2 \tau^2 L^2}.
\end{aligned}
\label{equation: accumulated updates}
\end{equation}
Before summing up (\ref{equation: accumulated updates}) for $n=1,2,\cdots, N$, it is noticed that $F(\bm\psi^{n, \tau}) \neq F(\bm\psi^{n+1, 0})$ on account of the quantization error. Their difference is thus taken into consideration as
\begin{equation}
\begin{aligned}
& F(\bm\psi^{n+1,0}) - F(\bm\psi^{n,\tau}) = F\left(\sum^U_{i=1} w_i^n Q(\bm\theta_i^{n,\tau})\right) - F(\bm\psi^{n,\tau}) \\
& \overset{(a)}{=} \Big\langle \nabla F(\bm\psi^{n,\tau}), \sum^U_{i=1} w_i^n \left(Q(\bm\theta_i^{n,\tau}) - \bm\theta_i^{n,\tau} \right) \Big\rangle + \frac{L}{2} \left\| \sum^U_{i=1} w_i^n \left(Q(\bm\theta_i^{n,\tau}) - \bm\theta_i^{n,\tau} \right)\right\|^2,
\end{aligned}
\label{equation: loss functions difference}
\end{equation}
where $(a)$ is due to \textbf{Assumption \ref{assumption: smooth}} and \textbf{Definition \ref{defintion}}. Based on independence of quantization results among clients and \textbf{Lemma \ref{lemma: quantization property}}, the expectation of (\ref{equation: loss functions difference}) is given by
\begin{equation}
\mathbb E\left[ F(\bm\psi^{n+1,0}) - F(\bm\psi^{n,\tau}) \right]  \leq \frac{L}{2} \sum^U_{i=1} w_i^n \frac{Z ( \theta_{i}^{n, \max})^2}{4(2^{q_i^n} - 1)^2}.
\label{equation: loss function difference expectation}
\end{equation}
Finally, we can sum up (\ref{equation: accumulated updates}) for $n = 1,2,\cdots, N$ and substitute (\ref{equation: loss function difference expectation}) into the sum formula, so that we complete the proof of \textbf{Theorem \ref{theorem: round}}.
\end{IEEEproof}
For $m=0$ of the left accumulated terms in \textbf{Theorem \ref{theorem: round}}, we have $\mathbb E[\|\nabla F(\bm \psi^{n,0})\|^2] = \mathbb E[\|\nabla F(\bm \theta^{n-1})\|^2]$. And $\mathbb E[\|\nabla F(\bm \psi^{n,m})\|^2]$ for $m\leq 1$ is also strongly tied to $\mathbb E[\|\nabla F(\bm \theta^{n-1})\|^2]$. Hence, the upper bound of global gradients can help to guarantee the FL convergence \cite{knob_choose}. It can be noticed that the upper bound in \textbf{Theorem \ref{theorem: round}} consists of 3 parts: the descent of the loss function, the quantization error, and the data property. Since $\bm\theta^0$ is initiated and $\bm \theta^N$ arrives at the optimal point, the first term is fixed. The second term implies that $\bm q^n$ and $\bm a^n$ jointly affect the quantization error. And later terms are all about the data property and affect the convergence by $\bm a^n$. As \textbf{Theorem \ref{theorem: round}} indicates, it is $\bm q^n$ and $\bm a^n$ that entirely dictate the performance of our FL.
\par
\section{Problem Formulation}
In this section, an optimization problem is formulated to minimize the energy consumption of clients. Latency and energy consumption in both communication and computation are modeled. And constraints about maximal latency and channels are given. Furthermore, the derived upper bound in Section \ref{section: convergence} also gives constraints to guarantee the convergence.
\subsection{Communication Model}
The the communication process in the $n$-th communication round contains the downlink communication and the uplink communication. In a real scenario, communication channels vary with time. But in one communication round, channel responses are supposed to be constant \cite{OFDMA}. And benefitting from channel estimation technique \cite{channel_estimation}, channel responses can be obtained.
\par
For the downlink communication, latency can be ignored compared to the uplink communication due to sufficient bandwidths and large transmitting power \cite{whole_optimization}.
The downlink energy consumption of the server is out of consideration, since we only concentrate on the energy consumption of clients.
\par
Different from the downlink communication, the uplink communication demands each participating client has a private channel and uploads its quantized local model on this channel. We suppose that the uplink communication is in the orthogonal frequency division multiple access (OFDMA) system with $C$ channels. Then, the allocation indicator vector $\bm{r}_i^n=[r_{i,1}^n, r_{i,2}^n, \cdots , r_{i,C}^n]^\top$ denotes the channel allocation result, where each indicator scalar $r_{i,c}^n\in \{0,1\}$ denotes the allocation state of channel $c$ to client $i$ with $r_{i,c}^n = 1$ indicating channel $c$ is assigned to client $i$ in the $n$-th communication round, and vice versa. In the OFDMA system, a channel should be allocated to no more than one client, i.e.,
\begin{equation}
\sum^U_{i=1} r_{i,c}^n \leq 1.
\label{equation: channel constraint}
\end{equation}
As for a participating client, it is essential to be assigned a channel to upload the quantized local model, so the constraint corresponding to the client is given by
\begin{equation}
\sum^C_{c=1} r_{i,c}^n = a_i^n.
\label{equation: participation constraint}
\end{equation}
All allocation indicator vectors $\bm{r}^n_1, \bm{r}^n_2, \cdots, \bm{r}^n_U$ are stacked into a $C \times U$ matrix $\bm{R}^n$. Furthermore, column $i$ helps to express the uplink rate of client $i$ as
$v_i^n= \sum^C_{c=1} r^{n}_{i,c} B \log_2(1 + \frac {p h_{i, c}^{n}} {B N_0}),$
where $B$ is the bandwidth of each channel, $p$ is the transmitting power, $h_{i, c}^{n}$ is uplink channel response of client $i$ on channel $c$, and $N_0$ is the noise power spectral density. $h_{i, c}^{n}$ includes device gain, large scale fading with respect to client $i$, and small scale fading with respect to channel $c$, i.e., $h^{n}_{i, c} = h^{\rm Gain}h^{n, {\rm Rician}}_{i, c} h^{n, {\rm Loss}}_i.$ $h^{\rm Gain}$ consists of antenna gain and gain of other settings. Considering frequency selective channels on account of the multipath effect, $h_{i,c}^{n, {\rm Rician}}$ of small scale fading follows a $(K, \zeta)$ Rician distribution. According to the scenario in \cite{channel_model}, large scale fading can be determined by the distance  $d_i$ between the server and client $i$ and the carrier frequency $\nu$.
\par
With the uplink rate $v_i^n$, latency of the uplink communication is given by
\begin{equation}
T^{n, {\rm com}}_i = \frac{\ell_i^n}{v^n_i} = \frac{Z (q^n_i + 1) + 32}{\sum^C_{c=1} r^n_{i,c} B \log_2(1 + \frac {p h_{i, c}^{n}} {B N_0})}.
\label{equation: uplink time}
\end{equation}
Moreover, the energy consumption of the uplink communication can be written by
\begin{equation}
E_i^{n, {\rm com}} = p T_i^{n, {\rm com}}.
\label{equation: uplink energy}
\end{equation}
\par
In (\ref{equation: uplink time}) and (\ref{equation: uplink energy}), it is noticed that quantization level vector $\bm q^n$ and the channel allocation matrix $\bm R^n$ directly affect latency and the energy consumption in the uplink communication. And the participation vector $\bm a^n$ even determines the existence of latency and the energy consumption.
\subsection{Computation Model}
In \cite{senior_Kang_JSAC}, the computation time of $\tau^{\rm e}$ local epochs on the local set $\mathcal D_i$ is given by
\begin{equation}
T_i^{n, {\rm cmp}} = \tau^{\rm e} \frac{\gamma D_i} {f_i^n},
\label{equation: computation time}
\end{equation}
where $\gamma$ is the number of CPU cycles computing a sample, and $f_i^{n}$ is the CPU frequency of client $i$ in the $n$-th communication round. And the corresponding computation energy consumption of $\tau^{\rm e}$ local epochs is
\begin{equation}
E_i^{n, {\rm cmp}} = \tau^{\rm e} \alpha \gamma D_i (f_i^n)^2,
\label{equation: computation energy}
\end{equation}
where $\alpha$ is the energy consumption coefficient, and CPU frequencies of all clients stack into $\bm f^n = [f_1^n, f_2^n, \cdots, f_U^n].$ Due to the hardware configuration, the CPU frequency is within the minimal frequency $T^{\min}$ to the maximal frequency $T^{\max}$ as
\begin{equation}
f^{\min} \leq f_i^n \leq f^{\max}.
\label{equation: CPU frequancey}
\end{equation}
\par
We notice that $D_i$ plays a role in (\ref{equation: computation time}) and (\ref{equation: computation energy}) so that different dataset sizes among clients lead to the divergence of latency and the energy consumption, then, the process of wireless FL is influenced. (\ref{equation: computation time}) and (\ref{equation: computation energy}) suggest that it is significant to choose the CPU frequency for client $i$, since a large $f_i^n$ leads to huge energy consumption and a small $f_i^n$ leads to too much latency.
\subsection{Optimization Problem}
For each participating client, the server cannot wait unlimitedly for its local updates and uplink communication. Hence, participating client $i$ has a maximum latency constraint as
\begin{equation}
a_i^n(T_i^{n, {\rm cmp}} + T_i^{n, {\rm com}}) \leq T^{\max}.
\label{equation: latency constraint}
\end{equation}
\par
Motivated by \cite{whole_optimization}, parts of the upper bound are regarded as constraints to guarantee FL performance. In (\ref{equation: upper bound round}), we detach parts of the quantization error and the data property, and limit them from tending too much, that is
\begin{equation}
\sum^{N-1}_{n=0}\sum_{i=1}^U \left( 4\tau (1 - a_i^n w_i) (G_i^n)^2 + A_1 w_i^n (G_i^n)^2 + A_2 w_i^n(\sigma_i^n)^2 \right) \leq \epsilon_1,
\label{equation: quantization error constraint}
\end{equation}
\begin{equation}
\frac{L}
{2}\sum\limits_{n = 0}^{N - 1} {\sum\limits_{i = 1}^U {w_i^n \frac{ Z (\theta_{i}^{n, \max})^2 }{4(2^{q_i^n} - 1)^2} }} \leq \epsilon_2,
\label{equation: data property constraint}
\end{equation}
where $A_1 = \frac{2\eta^2 L^2(2\tau^3-3\tau^2+\tau)} {3-6\eta^2L^2\tau^2}$ and $A_2 = \eta L\tau + \frac{\eta^2L^2 (\tau^2-\tau)}{1 - 2\eta^2L^2\tau^2}$.
\par
Under constraints of wireless channels, latency, and the convergence, it is essential to reduce the total energy consumption as much as possible. Hence, the energy consumption is set as the objective function and (\ref{equation: channel constraint}), (\ref{equation: participation constraint}) and (\ref{equation: CPU frequancey})$\sim$(\ref{equation: data property constraint}) serve as constraints. Denoting aforementioned variables by $\bm X^n = \{ \bm f^n, \bm q^n, \bm a^n, \bm R^n \}$, the optimization problem is
\begin{equation}
\begin{aligned}
& \textbf{P1:} \min_{\bm X^n} \lim_{N\rightarrow +\infty} \sum^{N-1}_{n=0} \sum_{i = 1}^U a_i^n( E_i^{n, {\rm cmp}} + E_i^{n, {\rm com}}), \\
{\rm s.t.} \quad & \textbf{C1:} \quad a_i^n,\thinspace r_{i,c}^n \in \{0, 1\}, \quad \textbf{C2:} \quad \sum^C_{c=1} r_{i,c}^n = a_i^n, \quad \textbf{C3:} \quad \sum^U_{i=1} r_{i,c}^n \leq 1, \\
& \textbf{C4:} \quad a_i^n(T_i^{n, {\rm cmp}} + T_i^{n, {\rm com}}) \leq T^{\max}, \quad \textbf{C5:} \quad f^{\min} \leq f_i^n \leq f^{\max}, \\
& \textbf{C6:} \quad \lim_{N\rightarrow +\infty} \frac{1}{N} \sum^{N-1}_{n=0}\sum_{i=1}^U \left( 4\tau (1 - a_i^n w_i) (G_i^n)^2 + A_1 w_i^n (G_i^n)^2 + A_2 w_i^n(\sigma_i^n)^2 \right) \leq \epsilon_1, \\
& \textbf{C7:} \quad \lim_{N\rightarrow +\infty} \frac{1}{N} \sum\limits_{n = 0}^{N - 1} {\sum\limits_{i = 1}^U {w_i^n \frac{ ZL (\theta_{i}^{n, \max})^2 }{8(2^{q_i^n} - 1)^2} }} \leq \epsilon_2, \quad \textbf{C8:} \quad q_i^n \in \mathbb N^+.
\end{aligned}
\label{equation: old problem}%
\end{equation}
where $N\rightarrow+\infty$ represents that FL finally achieves the convergence. \textbf{P1} is a long-term problem and channel responses can not be estimated as a priori. Besides, we notice \textbf{P1} is a mixed integer nonlinear program (MINLP), which needs an ingenious algorithm to avoid the high complexity.
\section{Problem Solution}
In this section, the long-term optimization problem is transformed into a one-communication-round optimization problem resorting to Lyapunov optimization. The one-communication-round problem is then divided into a continuous subproblem and a combinatorial subproblem, which can be solved by convex optimization methods and a genetic algorithm, respectively.
\subsection{Lyapunov Optimazation}
Motivated by \cite{Lyapunov_quantization}, the long-term constraints in \textbf{C6} and \textbf{C7} can be transformed into demands of queue stability by Lyapunov optimization \cite{Lyapunov_book}. Hence, we firstly introduce virtual queues corresponding to \textbf{C6} and \textbf{C7} as
\begin{equation}
\begin{aligned}
\lambda _1^{n + 1}  = \max \Big\{ \sum_{i = 1}^U & \left( 4\tau (1 - a_i^nw_i) \left( {G_i^n } \right)^2 + A_1 w_i^n \left( {G_i^n } \right)^2 + A_2 w_i^n \left( {\sigma _i^n } \right)^2  \right) + \lambda _1^n  - \epsilon _1 ,0 \Big\},
\end{aligned}
\label{equation: virtual queue scheduling}
\end{equation}
\begin{equation}
\lambda _2^{n + 1}  = \max \Big\{ {\sum\limits_{i = 1}^U {w_i^n \frac{{ZL\left( { \theta _i^{n,\max } } \right)^2}} {{8(2^{q_i^n }  - 1)^2 }}} + \lambda _2^n  - \epsilon _2,0} \Big\}.
\label{equation: virtual queue error}
\end{equation}
Satisfying \textbf{C6} and \textbf{C7} is equivalent to the mean-rate stability of $\lambda_1^n$ and $\lambda_2^n$, which can be expressed by
$\lim_{n\rightarrow + \infty} \frac{\mathbb E [ \lambda_1^n]}{n} = 0$ and $\lim_{n\rightarrow + \infty} \frac{\mathbb E [ \lambda_2^n]}{n} = 0.$
As a measure of the queue length, the Lyapunov drift function is defined by
$\Delta^n \triangleq \frac{1}{2} (\lambda_1^n)^2 + \frac{1}{2} (\lambda_2^n)^2.$
Taking the objective function in \textbf{P1} into consideration, the Lyapunov drift-plus-penalty function is defined by
\begin{equation}
\Delta_V^n \triangleq \mathbb E\big[\Delta^{n+1} - \Delta^n + V\sum\limits_{i = 1}^U a_i^n \left( { E_i^{n, {\rm cmp}}  + E_i^{n, {\rm com}} } \right) | \lambda_1^n, \lambda_2^n \big],
\label{equation: Lyapunov drift-plus-penalty}
\end{equation}
where $V$ is a penalty weight factor to make a trade-off between the energy consumption and FL performance. A large $V$ emphasizes reducing the energy consumption and neglects FL performance, and vice versa. Herein, $\Delta_V^n$ can serve as a new objective function and should be minimized. Due to the complex form of $\Delta_V^n$, we only keep cross terms similarly in \cite{Lyapunov_quantization} as
\begin{equation}
\begin{aligned}
\Delta_V^n \leq & \left( {\lambda _1^n  - \epsilon _1 } \right) \sum\limits_{i = 1}^U {\left( {4\tau (1 - a_i^nw_i) \left( {G_i^n } \right)^2  + A_1 w_i^n \left( {G_i^n } \right)^2  + A_2 w_i^n \left( {\sigma _i^n } \right)^2 } \right)}   \\
& + \left( {\lambda _2^n  - \epsilon _2 } \right)\sum\limits_{i = 1}^U {w_i^n \frac{{ZL\left( {\theta _i^{n,\max } } \right)^2}}
{{8(2^{q_i^n }  - 1)^2 }}} + A_0  + V\sum\limits_{i = 1}^U {a_i^n \left( { E_i^{n, {\rm cmp}}  + E_i^{n, {\rm com}} } \right)},
\end{aligned}
\label{equation: Lyapunov upper bound}
\end{equation}
where $A_0$ is a positive constant.
Ignoring $A_0$, the new optimization problem in the $n$-th communication round is
\begin{equation}
\textbf{P2:} \min_{\bm X^n} J^n, \qquad \textbf{s.t.} \quad \textbf{C1} \sim \textbf{C5} \enspace {\rm and} \enspace \textbf{C8}, \\
\label{equation: new optimization problem}
\end{equation}
where $J^n \triangleq \left( {\lambda _1^n  - \epsilon _1 } \right) \sum_{i = 1}^U {( {4\tau (1 - a_i^nw_i) \left( {G_i^n } \right)^2  + A_1 w_i^n \left( {G_i^n } \right)^2  + A_2 w_i^n \left( {\sigma _i^n } \right)^2 } )} + \left( {\lambda _2^n  - \epsilon _2 } \right) \times \sum\limits_{i = 1}^U {w_i^n \frac{{ZL\left( {\theta _i^{n,\max } } \right)^2}} {{8(2^{q_i^n }  - 1)^2 }}} + V\sum_{i = 1}^U {a_i^n \left( { E_i^{n,{\rm cmp}}  + E_i^{n, {\rm com}} } \right)} $. For the ($n$+1)-th communication round, a new optimization problem about $\bm X^{n+1}$ and $J^{n+1}$ can be formulated similarly. As such, the long-term problem \textbf{P1} is transformed into a one-communication-round problem \textbf{P2}. However, \text{P2} is also a MINLP and an improper solution procedure means great complexity. Herein, we present a low-complexity solution for \textbf{P2} in the sequel.
\subsection{Problem Decomposition}
Variables of $\bm X^n = \{ \bm f^n, \bm q^n, \bm a^n, \bm R^n \}$ can be categorized into two parts: $\bm a^n, \bm R^n$ are combinatorial variables, and $\bm f^n, \bm q^n$ are continuous variables. It is remarkable that $\bm q^n$ is an integer variable and can be relaxed into a continuous variable. Based on the two categories of variables, \textbf{P2} can be decomposed into two subproblems. According to Tammer decomposition \cite{Tammer}, \textbf{P2} is firstly transformed into an equivalent problem as
\begin{equation}
\textbf{P3:} \min_{\bm a^n, \bm R^n} \left(\min_{\bm f^n, \bm q^n} J^n(\bm f^n, \bm q^n, \bm a^n, \bm R^n) \right), \qquad \textbf{s.t.} \quad \textbf{C1} \sim \textbf{C5} \enspace {\rm and} \enspace \textbf{C8}.
\label{equation: master optimization problem}
\end{equation}
\par
The optimization problem $\textbf{P3}$ consists of an outer subproblem about $\bm a^n, \bm R^n$ and an inner subproblem about $\bm f^n, \bm q^n$. Specifically, the outer optimization subproblem is
\begin{equation}
\begin{aligned}
\textbf{P3.1:} \min_{\bm a^n , \bm R^n} J_1^n(\bm a^n, \bm R^n)  \qquad \textbf{ s.t.} \enspace \textbf{C1} \sim \textbf{C4},
\end{aligned}
\label{equation: outer optimization problem}
\end{equation}
where $J_1^n(\bm a^n, \bm R^n) = J^n(\bm f^{n*}, \bm q^{n*}, \bm a^n, \bm R^n)$ and $(\bm f^{n*}, \bm q^{n*})$ is the optimal point of the inner optimization subproblem. With fixed $(\bm a^n, \bm R^n)$, the inner subproblem is written by
\begin{equation}
\begin{aligned}
\textbf{P3.2:} \min_{\bm f^n , \bm q^n} J_2^n(\bm f^n , \bm q^n)  \qquad \textbf{s.t.} \enspace \textbf{C4}, \textbf{C5} \enspace {\rm and} \enspace \textbf{C8},
\end{aligned}
\label{equation: inner optimization problem}
\end{equation}
where $J_2^n(\bm f^n , \bm q^n) =  \left( {\lambda _2^n  - \epsilon _2 } \right) \enspace \sum_{i = 1}^U {w_i^n \frac{{ZL\left( { \theta _i^{n,\max } } \right)^2}}{{8(2^{q_i^n }  - 1)^2 }}} + V\sum_{i = 1}^U a_i^n \Big( \tau^{\rm e} \alpha \gamma D_i (f_i^n)^2  + \frac{p(Zq_i^n + Z +32)}{v_i^n} \Big)$. In $J_2^n$, terms unrelated to $\bm f^n , \bm q^n$ in $J^n$ are ignored. Thanks to the sum form of $J_2^n$ and individual constraints among clients, \textbf{P3.2} can be further decomposed into a series of individual optimization problems corresponding to each participating client $i \in \mathcal U^n$ as
\begin{equation}
\textbf{P3.2}'\textbf{:} \min_{f^n_i , q^n_i} J_3^n(f_i^n, q_i^n) \qquad \textbf{s.t.} \enspace \textbf{C5}, \textbf{C8} \enspace {\rm and} \enspace \textbf{C4}'\textbf{:} \enspace T_i^{n, {\rm cmp}} + T_i^{n, {\rm com}} \leq T^{\max},
\label{equation: inner optimization split problem}
\end{equation}
where $J_3^n(f_i^n, q_i^n) = \left( {\lambda _2^n  - \epsilon _2 } \right) {w_i^n \frac{{ZL\left( { \theta _i^{n,\max } } \right)^2}}{{8(2^{q_i^n }  - 1)^2 }}} + V \tau^{\rm e} \alpha \gamma D_i (f_i^n)^2 + \frac{pVZq_i^n}{v_i^n}$. Relaxing \textbf{C8} to a continuous constraint, $\textbf{P3.2}'$ is thus transformed into a continuous optimization problem as
\begin{equation}
\textbf{P3.2}''\textbf{:} \min_{f^n_i , q^n_i} J_3^n(f_i^n, q_i^n) \quad \textbf{ s.t.} \enspace \textbf{C4}', \textbf{C5} \enspace {\rm and} \enspace \textbf{C8}'\textbf{:} \ q_i^n \geq 1.
\label{equation: inner optimization relaxation problem}
\end{equation}
In fact, $(\hat f_i^{n*}, \hat q_i^{n*})$, the optimal point of $\textbf{P3.2}''$ hardly satisfies \textbf{C8}. However, $q_i^{n*}$ can be indeed computed by $\hat q_i^{n*}$, which will be introduced in detail in Section V-C.
\par
After the above transforms of problems, we decompose the optimization problem \textbf{P2} into a combinatorial optimization subproblem \textbf{P3.1} and a continuous optimization subproblem $\textbf{P3.2}''$. In Section V-C and Section V-D, these two subproblems will be solved respectively.
\subsection{Continuous Optimization Subproblem}
With computing $\frac{\partial^2 J_3^n}{\partial (f_i^n)^2} > 0$, $\frac{\partial^2 J_3^n}{\partial (q_i^n)^2} > 0$ and $\frac{\partial^2 J_3^n}{\partial f_i^n \partial q_i^n} = \frac{\partial^2 J_3^n}{\partial q_i^n \partial f_i^n} = 0$, it is can be proved that $J_3^n$ is convex. In addition, it is easy to prove that $\textbf{C4}', \textbf{C5}, \textbf{C8}'$ are convex sets when $\lambda_2^n - \epsilon > 0$. Hence, $\textbf{P3.2}''$ is a convex optimization problem. With the help of the Lagrange multipliers technique, a closed-form solution can be obtained. Firstly, the Lagrange function is expressed by
$
\Lambda = \left( {\lambda _2^n  - \epsilon _2 } \right) {w_i^n \frac{{ZL\left( { \theta _i^{n,\max } } \right)^2}}{{8(2^{q_i^n }  - 1)^2 }}} + V \tau^{\rm e} \alpha \gamma D_i (f_i^n)^2 + \frac{pVZq_i^n}{v_i^n} + \kappa_1 \left(\tau^{\rm e} \frac{\gamma D_i} {f_i^{n}} + \frac{Zq_i^{n} + Z + 32}{v_i^n} - T^{\max}\right) + \kappa_2 (f^{\min} - f_i^n) + \kappa_3 (f_i^n - f^{\max}) + \kappa_4 (1 - q_i^n).
$
Thus, KKT conditions are given as
\begin{small}
\begin{equation}
\begin{cases}
\tau^{\rm e} \frac{\gamma D_i} {f_i^{n}} + \frac{Zq_i^{n} + Z + 32}{v_i^n} \leq T^{\max}, \quad f^{\min} \leq f_i^{n} \leq f^{\max}, \quad q_i^n \geq 1, \quad \kappa_1, \kappa_2, \kappa_3, \kappa_4 \geq 0, \\
\kappa_1 \Big(\tau^{\rm e} \frac{\gamma D_i} {f_i^{n}} + \frac{Zq_i^{n} + Z + 32}{v_i^n} - T^{\max}\Big) = 0, \enspace \kappa_2(f^{\min} - f_i^{n}) = 0, \enspace \kappa_3(f_i^{n} - f^{\max}) = 0, \enspace \kappa_4(1-q_i^n) = 0,\\
2V\tau^{\rm e} \alpha \gamma D_i f_i^n - \kappa_1 \tau^{\rm e} \frac{\gamma D_i}{(f_i^{n})^2} - \kappa_2 + \kappa_3 = 0, \quad \frac{pV}{v_i^n} - 2^{q_i^n} \ln2 \cdot \left( {\lambda _2^n  - \epsilon _2 } \right) {w_i^n \frac{{L\left( { \theta _i^{n,\max } } \right)^2}}{{4(2^{q_i^n }  - 1)^3 }}} + \frac{\kappa_1}{v_i^n} - \frac{\kappa_4}{Z}= 0.
\end{cases}
\label{equation: KKT condition}
\end{equation}
\end{small}%
Despite (\ref{equation: KKT condition}) existing, obtaining the optimal point is not immediately clear. Consequently, we divide $\textbf{P3.2}''$ into 5 cases and solve KKT conditions in each case. These cases are listed as follows. \emph{Case 1: } $\textbf{C8}'$ is strict. \emph{Case 2:} $\textbf{C8}'$ is loose and $\textbf{C4}'$ is loose. \emph{Case 3:} $\textbf{C8}'$ is loose, $\textbf{C4}'$ is strict and the maximum in $\textbf{C5}$ is gotten. \emph{Case 4:} $\textbf{C8}'$ is loose, $\textbf{C4}'$ is strict and the minimum in $\textbf{C5}$ is gotten. \emph{Case 5:} $\textbf{C8}'$ is loose, $\textbf{C4}'$ is strict and $\textbf{C5}$ is loose.
\par
We notice that the five cases are complete and any two cases are mutually exclusive. Furthermore, a relationship between $\textbf{C4}'$ and $\textbf{C5}$ also helps to solve KKT conditions as follows.
\begin{lemma}
If $\textbf{C4}'$ of $\textbf{P3.2}''$ is loose for the optimal point $(\hat f_i^{n*}, \hat q_i^{n*})$, $\hat f^{n*}_i$ gets its minimum $f^{\min}$ in $\textbf{C5}$.
\label{lemma: loose strict}
\end{lemma}
\begin{IEEEproof}
With the proof by contradiction, we firstly suppose $\hat f^{n*}_i > f^{\min}$. And we have $\tau^{\rm e} \frac{\gamma D_i} {\hat f_i^{n*}} + \frac{Z\hat q_i^{n} + Z + 32}{v_i^n} < T^{\max}$ since $\textbf{C4}'$ is loose. Hence, there is a frequency satisfying $\tilde f = \max \left\{ f^{\min}, \frac{v_i^n \tau^{\rm e} \gamma D_i}{v_i^n T^{\max} - Z \hat q_i^n - Z - 32} \right\}$. It is obvious that $\tilde f < \hat f_i^{n*}$ and $(\tilde f, \hat q_i^{n*})$ satisfies $\textbf{C4}', \textbf{C5}$ and $\textbf{C8}'$. Moreover, $J_3^n$ is a strictly increasing function of $f_i^n$ according to $\frac{\partial J_3^n}{\partial f_i^n} > 0$. $J_3^n(\tilde f, \hat q_i^{n*}) < J_3^n(\hat f_i^{n*}, \hat q_i^{n*})$ is then proved, which is contradictory to the optimality of $(\hat f_i^{n*}, \hat q_i^{n*})$. As thus, the initial assumption $\hat f^{n*}_i > f^{\min}$ fails and we complete the proof of \textbf{Lemma \ref{lemma: loose strict}}.
\end{IEEEproof}
\textbf{Lemma \ref{lemma: loose strict}} helps to solve (\ref{equation: KKT condition}) in \emph{Case 2}. Denoting the optimal point of case $j$ by $(\hat f_i^{n*_j}, \hat q_i^{n*_j})$, solutions and corresponding prerequisites to check in all cases are given as follows.
\par
\emph{Case 1:} With $\hat q_i^{n*_1} = 1$, there is only $f_i^n$ to solve. And $\frac{\partial J_3^n}{\partial f_i^n} > 0$ means that a lower $f_i^n$ leads to a less $J_3^n$. Constrained by $\textbf{C4}'$ and $\textbf{C5}$, $\hat f_i^{n*}$ is computed by $\hat f_i^{n*_1} = \max \left\{ f^{\min}, \frac{v_i^n \tau^{\rm e} \gamma D_i}{v_i^n T^{\max} - 2 Z - 32} \right\}.$ And $(\hat f_i^{n*_1}, \hat q_i^{n*_1})$ should satisfy (\ref{equation: KKT condition}), where the only prerequisite needed to check is
\begin{equation}\textbf{Pre1:}\quad pV -  \frac{1}{2}v_i^nw_i^n L (\lambda_2^n - \epsilon_2)  (\theta_i^{n, \max})^2 \ln2 \geq 0.\end{equation}
\par
\emph{Case 2:} Based on \textbf{Lemma \ref{lemma: loose strict}}, we have $\hat f_i^{n*_2} = f^{\min}$. With $\kappa_1=\kappa_4=0$, we can utilize the formula of roots of a cubic equation without the quadratic term to solve $\hat q_i^n$ from (\ref{equation: KKT condition}) as $\hat q_i^{n*_2} = \log_2[1+ \sqrt[3] A_4( \sqrt[3]{\frac{1}{2} + \sqrt{\frac{1}{4} - \frac{A_4}{27}}} + \sqrt[3]{\frac{1}{2} - \sqrt{\frac{1}{4} - \frac{A_4}{27}}})]$, where $A_4 = \frac{v_i^n w_i^n L (\lambda_2^n - \epsilon_2) (\theta_i^{n, \max})^ 2 \ln 2}{4pV}$. And residual prerequisites needed to check in (\ref{equation: KKT condition}) are
\begin{equation}\textbf{Pre2:}\quad \tau^{\rm e} \frac{\gamma D_i} {f^{\min}} + \frac{Zq_i^{n} + Z + 32}{v_i^n} < T^{\max}, \  {\rm and} \ \hat q_i^{n*_2} > 1.\end{equation}
\par
\emph{Case 3:} It is easy to solve $\hat f_i^{n*_3} = f^{\max}$ and $\hat q_i^{n*_3} = \frac{ f^{\max} v_i^n T^{\max} - v_i^n \tau^{\rm e} \gamma D_i - f^{\max}(Z+32)}{f^{\max} Z}$. And we have 3 prerequisites in (\ref{equation: KKT condition}) to check, that is
\begin{equation}\textbf{Pre3:} \quad  \kappa_1 = v_i^n w_i^n L (\lambda_2^n - \epsilon_2) \frac{ (\theta_i^{n, \max})^2 2^{q_i^{n*_3}}}{4(2^{q_i^{n*_3}} - 1)^3} \ln 2 \geq pV, \ 2V \alpha  (f^{\max})^3 \leq \kappa_1, \ {\rm and} \ \hat q_i^{n*_3} > 1.\end{equation}
\par
\emph{Case 4:} Similarly to \emph{Case 3}, we solve $\hat f_i^{n*_4} = f^{\min}$ and $\hat q_i^{n*_4} = \frac{ f^{\min} v_i^n T^{\max} - v_i^n \tau^{\rm e} \gamma D_i - f^{\min}(Z+32)}{f^{\min} Z}$. Prerequisites in (\ref{equation: KKT condition}) to check are
\begin{equation}\textbf{Pre4:}\quad \kappa_1 = v_i^n w_i^n L (\lambda_2^n - \epsilon_2) \frac{ (\theta_i^{n, \max})^2 2^{q_i^{n*_4}}}{4(2^{q_i^{n*_4}} - 1)^3} \ln 2 \geq pV, \ 2V \alpha (f^{\min})^3 \geq \kappa_1, \ {\rm and} \ \hat q_i^{n*_4} > 1.\end{equation}
\par
\emph{Case 5:} Without any strict constraints of $f_i^n$ or $q_i^n$, the optimal point can not directly be computed. Nevertheless, $\hat f_i^{n*_5}$ can be expressed by $\hat q_i^{n*_5}$ since $\textbf{C4}'$ is strict. Substituting $\hat f_i^{n*_5} = \frac{v_i^n \tau^{\rm e} \gamma D_i}{v_i^n T^{\max} - Z \hat q^{n*_5}_i - Z - 32}$ into (\ref{equation: KKT condition}), we have a transcendental equation of $\hat q^{n*_5}_i$ as
\begin{equation}
p + 2\alpha \left( \frac{v_i^n \tau^{\rm e} \gamma D_i}{v_i^n T^{\max} - Z\hat q_i^{n*_5} - Z - 32} \right)^3  = v_i^n w_i^n L (\lambda_2^n - \epsilon_2) \frac{(\theta_i^{n, \max})^2 2^{\hat q_i^{n*_5}}}{4V(2^{\hat q_i^{n*_5}} - 1)^3} \ln 2.
\label{equation: transcendental equation}
\end{equation}
There is no closed-form solution for (\ref{equation: transcendental equation}), however, an approximate analytical solution can be obtained by the use of first-order Taylor expansion on $q_i^{n'*}$, where $n'$ is the index of the communication round when client $i$ last participates. Due to small changes of models within several communication rounds, $q_i^{n'*}$ is close to $q_i^{n*5}$ and the Taylor expansion on $q_i^{n'*}$ can well approximate (\ref{equation: transcendental equation}). Hence, we have the approximate solution as
\begin{equation}
\hat q_i^{n*_5} = q_i^{n'*} + \frac{v_i^n w_i^n L (\lambda_2^n - \epsilon_2) \frac{(\theta_i^{n, \max})^2 2^{\hat q_i^{n'*}}}{4V(2^{\hat q_i^{n'*}} - 1)^3} \ln 2 - 2\alpha \left( \frac{v_i^n \tau^{\rm e} \gamma D_i}{v_i^n T^{\max} - Z q_i^{n'*} - Z - 32} \right)^3 - p} {v_i^n w_i^n L (\lambda_2^n -\epsilon_2) \frac{(\theta_i^{n, \max})^2 (2\cdot 2^{2q_i^{n'*}} + 1) 2^{q_i^{n'*}}}{4V(2^{q_i^{n'*}} - 1)^4} \ln¡­^2 2 + \frac{6\alpha Z (v_i^n \tau^{\rm e} \gamma D_i)^3}{(v_i^nT^{\max} - Z q_i^{n'*} - Z - 32)^4}}.
\label{equation: approximate solution}
\end{equation}
There are 2 prerequisites to check as \begin{equation}\textbf{Pre5:} \quad f^{\min} < \hat f_i^{n*_5} < f^{\max}, \ {\rm and }\ q^{n*_5}_i > 1.\end{equation}
\par
It is noticed that only in \emph{Case 5} does neither $\hat q_i^{n*}$ nor $\hat f_i^{n*}$ get boundary values. For most normal channel responses and suitable dataset sizes, in fact, the optimal point is away from boundary values. Therefore, (\ref{equation: approximate solution}), the closed-form solution of \emph{Case 5} is often optimal and can give some insights about choosing the quantization level. On account of $q_i^n$ varying with communication rounds and clients, we focus on the influence of $n$ and $D_i$.
\par
To simplify (\ref{equation: approximate solution}), we suppose that differences of $v_i^n, \theta_i^{n, \max}$ and $w_i^n$ can be omitted in different communication rounds. The numerator replacing $n$ with $n'$ in (\ref{equation: approximate solution}) is equal to zero since $q_i^{n'*}$ is optimal in the $n'$-th communication round and satisfies (\ref{equation: transcendental equation}). As for the original numerator, the coefficient of the positive term gets large due to $\lambda_2^n$ keeping rising before equilibrium. Consequently, the numerator in (\ref{equation: approximate solution}) is positive and we have $\hat q^{n*_5}_i > q_i^{n'*}$, which reflects the rising trend of $\hat q^{n*_5}_i$.
\par
As for different clients, we also suppose differences of $v_i^n$ and $\theta_i^{n, \max}$ can be omitted. $w_i^n$ can not be ignored due to $w_i^n = \frac{D_i}{\sum_{j \in \mathcal U^n} D_j}$. Focusing on $D_i$, (\ref{equation: approximate solution}) is rewritten by $\hat q_i^{n*_5} = q_i^{n'*} - C_0 + \frac{C_1}{C_2 D_i^2 + C_3},$ where $C_0, C_2, C_3$ are positive terms and $C_1 = v_i^n L (\lambda_2^n - \epsilon_2) \frac{(\theta_i^{n, \max})^2 2^{q^{n'*}_i}}{4V(2^{q^{n'*}_i} - 1)^3} \ln2 - p + C_0C_3$. Since $\lambda_2^n$ keeps rising and $2^{q^{n'*}_i}$ is optimal, we prove $C_1>0$, deducing $D_i < D_j \Rightarrow q^{n*_5}_i > q^{n*_5}_j$.
\par
To sum up the above derivation and the analysis, we have two remarks as follows.
\par
\emph{\textbf{Remark 1:} $\hat q_i^{n*_5}$ rises with $n$, and tends to a stationary point when $\lambda_2^n$ reaches equilibrium.}
\par
\emph{\textbf{Remark 2:} $\hat q_i^{n*_5}$ is negatively correlated with $D_i$, i.e., a client with a large dataset tends to adopt a low quantization level and vice versa.} \par
If varying channel responses are neglected, the two remarks are consistent with demands in wireless FL. Specifically, a low quantization level leads to a high quantization error which has a small impact due to inherent fluctuation in early communication rounds. When the model tends to the convergence, a high quantization level reduces the quantization error to avoid fluctuation. Under communication constraints, computation on a large dataset takes much latency, and the client decreases quantization levels to shorten communication latency for compensation.
\par
In the end, a united formula of $\hat q_i^{n*}$ is given by
\begin{equation}
\hat q_i^{n*} =
\begin{cases}
1, \quad &{\rm if\ \textbf{Pre1}\ is \ satisfied};\\
\log_2[1+ \sqrt[3] A_4( \sqrt[3]{\frac{1}{2} + \sqrt{\frac{1}{4} - \frac{A_4}{27}}} + \sqrt[3]{\frac{1}{2} - \sqrt{\frac{1}{4} - \frac{A_4}{27}}})], \quad &{\rm if\ \textbf{Pre2}\ is \ satisfied};\\
\frac{  v_i^n T^{\max} - \frac{v_i^n \tau^{\rm e} \gamma D_i}{f^{\max}} - Z -32}{ Z}, \quad &{\rm if\ \textbf{Pre3}\ is \ satisfied};\\
\frac{  v_i^n T^{\max} - \frac{v_i^n \tau^{\rm e} \gamma D_i}{f^{\min}} - Z -32}{ Z}, \quad &{\rm if\ \textbf{Pre4}\ is \ satisfied};\\
\hat q_i^{n*_5}, \quad &{\rm if\ \textbf{Pre5}\ is \ satisfied}.
\end{cases}
\label{equation: solution of quantization}
\end{equation}
As we have mentioned in Section V-B, it is $q_i^{n*}$ rather than $\hat q_i^{n*}$ that satisfies \textbf{C8}. Fortunately, $(q_i^{n*}, f_i^{n*})$ can be derived by $\hat q_i^{n*}$ thanks to \textbf{Theorem \ref{theorem: solution}} as follows.
\begin{theorem}
$\textbf{S1}$ and $\textbf{S2}$ are defined by
$\textbf{S1:} \min_{x,y} W(x,y), \textbf{s.t.:}(x,y) \in \mathcal W,$
$\textbf{S2:} \min_{x,y} W(x,y),$ $ \textbf{s.t.} (x,y) \in \mathcal W, x \in \mathbb Z,$
where $W(x,y)$ is a convex function and $\mathcal W$ is a convex set. $\mathscr S(x)$ denotes the optimal solution of $y$ in \textbf{S1} for given $x$. If $(\hat x^*, \hat y^*)$ is the optimal point in $\textbf{S1}$ and $\textbf{S2}$ is solvable, the optimal point in $\textbf{S2}$ is either $(\lceil \hat x^* \rceil, \mathscr{S}(\lceil\hat x^*\rceil))$ or $(\lfloor \hat x^* \rfloor, \mathscr{S}(\lfloor\hat x^*\rfloor))$.
\label{theorem: solution}
\end{theorem}
\begin{IEEEproof}
For any point $(x_0, y_0) \in \mathcal W, x_0 \in \mathbb Z$, there are four relationships between $x_0$ and $\hat x^*$: $x_0 < \lfloor \hat x^* \rfloor$, $x_0 > \lceil \hat x^* \rceil$, $x_0 = \lfloor \hat x^* \rfloor$ or $x_0 = \lceil \hat x^* \rceil$. If $x_0 = \lfloor \hat x^* \rfloor$ or $x_0 = \lceil \hat x^* \rceil$, we have $\min \{W(\lfloor \hat x^* \rfloor, \mathscr S(\lfloor \hat x^* \rfloor)), W(\lceil \hat x^* \rceil, \mathscr S(\lceil \hat x^* \rceil)) \} \leq W(x_0, y_0)$ due to the definition of $\mathscr S(\cdot)$. For $x_0 > \lceil \hat x^* \rceil$, $\lceil \hat x^* \rceil$ can be expressed by $\lceil \hat x^* \rceil = \delta_1 \hat x^* + (1-\delta_1) x_0,\enspace \delta_1 \in (0, 1]$. And we have $(\lceil \hat x^* \rceil, y_1) \in \mathcal W$ for $y_1 = \delta_1 \hat y^* + (1-\delta_1) y_0$, since $\mathcal W$ is a convex set. According to the convexity of $W(x,y)$ and the optimality of $(\hat x^*, \hat y^*)$, we have $W(\lceil \hat x^* \rceil, y_1) \leq \delta_1 W(\hat x^*, \hat y^*) + (1-\delta_1) W(x_0, y_0) \leq W(x_0, y_0)$. Moreover, the definition of $\mathscr S(\cdot)$ tells $W(\lceil \hat x^* \rceil , \mathscr S(\lceil \hat x^* \rceil)) \leq W(\lceil \hat x^* \rceil, y_1)$, which deduces $W(\lceil \hat x^* \rceil , \mathscr S(\lceil \hat x^* \rceil)) \leq W(x_0, y_0)$. As for $x_0 < \lfloor \hat x^* \rfloor$, similarly, we have $W(\lfloor \hat x^* \rfloor, \mathscr S(\lfloor \hat x^* \rfloor)) \leq W(\lfloor \hat x^* \rfloor, y_2) \leq \delta_2 W(\hat x^*, \hat y^*) + (1-\delta_2) W(x_0, y_0) \leq W(x_0, y_0)$. To sum up, we can prove $\min \{W(\lfloor \hat x^* \rfloor, \mathscr S(\lfloor \hat x^* \rfloor)), W(\lceil \hat x^* \rceil, \mathscr S(\lceil \hat x^* \rceil)) \} \leq W(x_0, y_0)$ for any point $(x_0, y_0) \in \mathcal W, x_0 \in \mathbb Z$. This completes the proof of the optimality of $(\lceil \hat x^* \rceil, \mathscr{S}(\lceil\hat x^*\rceil))$ or $(\lfloor \hat x^* \rfloor, \mathscr{S}(\lfloor\hat x^*\rfloor))$.
\end{IEEEproof}
To simplify the form, we also give the definition $\mathscr{S}(q_i^{n}) \triangleq \max \{ f^{\min}, \frac{v_i^n \tau^{\rm e} \gamma D_i}{v_i^n T^{\max} - Z q_i^n - Z - 32} \}$. Similarly to \emph{Case 1},  $\mathscr{S}(q_i^n)$ is the optimal frequency in $\textbf{P3.2}'$ for fixed $q_i^n$. Based on \textbf{Theorem \ref{theorem: solution}}, we deduce that the optimal point in $\textbf{P3.2}'$ is
\begin{equation}
(q_i^{n*}, f_i^{n*}) = \label{equation: continuous solution}\begin{cases}
(\lfloor \hat q_i^{n*} \rfloor, \mathscr S(\lfloor \hat q_i^{n*} \rfloor)), \quad {\rm if\ } J_3^n(\lfloor \hat q_i^{n*} \rfloor, \mathscr S(\lfloor \hat q_i^{n*} \rfloor)) \leq J_3^n(\lceil \hat q_i^{n*} \rceil, \mathscr S(\lceil \hat q_i^{n*} \rceil)); \\
(\lceil \hat q_i^{n*} \rceil, \mathscr S(\lceil \hat q_i^{n*} \rceil)), \quad {\rm if\ } J_3^n(\lfloor \hat q_i^{n*} \rfloor, \mathscr S(\lfloor \hat q_i^{n*} \rfloor)) > J_3^n(\lceil \hat q_i^{n*} \rceil, \mathscr S(\lceil \hat q_i^{n*} \rceil)).
\end{cases}
\end{equation}
Finally, a series of optimal points of all participating clients in $\textbf{P3.2}'$ is the optimal point in \textbf{P3.2}.
\subsection{Combinatorial Optimization Subproblem}
Despite obtaining the closed-form solution of \textbf{P3.2}, it is difficult to solve optimal combinatorial variables $\bm a^{n*}$ and $\bm R^{n*}$ due to the nonlinear form of $J^n_1$ in \textbf{P3.1}. Motivated by \cite{genetic_algorithm}, a genetic algorithm can help to allocate channels in the OFDMA system. And \textbf{C2} can help to express $\bm a^{n*}$ by $\bm R^{n*}$. In the genetic algorithm, any channel allocation matrix is decoded into a chromosome. To simplify the formula, the superscript $n$ is omitted in the current section and $\bm R_s$ denotes a chromosome in the $s$-th generation $\mathcal R_s$. For each $\bm R_s \in \mathcal R_s$, its fitness function is computed by
\begin{equation}
J_4(\bm R_s) = (J_0^{\max} - J_0(\bm a_s, \bm R_s, \bm f_s^*, \bm q_s^*))^\iota,
\label{equation: fitness function}
\end{equation}
where $J_0^{\max}$ is the maximal $J_0$ among $\mathcal R_s$, and $\iota > 0$ is a exponential coefficient which adjusts the dispersion of the fitness function. We set $J_4(\bm R_s) = 0$ for infeasible $\bm R_s$. According to the fitness functions, $R_s^{\rm p}$ is selected to compose the $s$-th parent generation $\mathcal R_s^{\rm p}$. Thus, crossover on each pair of parent chromosomes and mutation on children chromosomes generate the $(s+1)$-th generation of chromosomes. The same steps are executed in the $(s+1)$-th generation and so on. The detailed process is presented in \textbf{Algorithm \ref{algorithm: genetic algorithm}}.
\begin{algorithm}
  \caption{Genetic Algorithm about Channel Allocation}
  \label{algorithm: genetic algorithm}
  \KwOut{optimal channel allocation matrix $\bm{R}^{*}$, optimal participation vector $\bm{a}^{*}$, optimal quantization level vector $\bm q^{*}$, optimal CPU frequency vector $\bm f^{*}$}
  Set maximal generation number $s_{\max}$, population $N^{\rm pop}$, crossover probability $p^{\rm c}$ and mutation probability $p^{\rm m}$, and generate the initial generation $\mathcal R_0$ randomly\;
  \While{$s < s_{\max}$}
  {
    Solve $\bm a_s, \bm q_s^*, \bm f_s^*$ with \textbf{C2} and (\ref{equation: continuous solution}), and then compute $J_4(\bm R_s)$ for each $\bm R_s \in \mathcal R_s$\;
    Select chromosomes with fitness functions to compose the $s$-th parent generation $\mathcal R_s^{\rm p}$\;
    \For{each pair in $\{(\bm R^{\rm p1}_s, \bm R^{\rm p2}_s) | \textit{ sample}\ \bm R^{\rm p1}_s, \bm R^{\rm p2}_s \in \mathcal R_s^{\rm p} \textit{\ without\ replacement}\}$}
    {
        Create children chromosomes $\bm R^{\rm c1}_s, \bm R^{\rm c2}_s$ with random crossover on $\bm R^{\rm p1}, \bm R^{\rm p2}$\;
        Mutate $\bm R^{\rm c1}_s, \bm R^{\rm c2}_s$ to generate $\bm R^{\rm c1'}_s, \bm R^{\rm c2'}_s$\;
        Add $\bm R^{\rm c1'}_s, \bm R^{\rm c2'}_s$ into the next generation $\mathcal R_{s+1}$\;
    }
    Update $s:= s+1$\;
  }
  Search for the best chromosome, i.e., $ \bm R^{*} = \mathop{\arg\max}_{\bm R_{s_{\max}} \in \mathcal R_{s_{\max}}} J_4(\bm R_{s_{\max}})$\;
  Solve $\bm a^{*}, \bm q^{*}, \bm f^{*}$ with \textbf{C2} and (\ref{equation: continuous solution}), and return $(\bm R^*, \bm a^*, \bm q^{*}, \bm f^{*})$.
\end{algorithm}

\section{Experiment Results}
To assess the energy consumption and FL performance of the QCCF algorithm we propose, our FL tasks are set in a circular network area including a server and ten clients which are uniformly distributed in the 500-meter-radius circular area. Other communication parameters and computation parameters are listed in Table \ref{table: parameter value}.
\par
\textbf{Datasets.} There are two FL tasks to validate our QCCF algorithm. One task is the handwritten digit and letter identification on the FEMNIST dataset from LEAF \cite{femnist}. Another is the colored image identification on the CIFAR-10 dataset \cite{cifar10}. More generally, we assume data owned by clients is non-independent and identically distributed. The dataset size follows a Gaussian distribution, i.e., $D_i \sim N(\mu, \beta)$. We set $\mu = 1200$ and $\beta = 150, 300$ in the following experiments.
\par
\textbf{Models.} Convolutional neural networks (CNNs) stand as effective tools for image identification tasks. For the first task on the FEMNIST dataset, we employ a CNN composed of two convolutional layers with 32 $1\times5\times5$ kernels and 64 $32\times5\times5$ kernels, respectively, and a hidden layer with 3136 neurons. Another CNN for the second task on the CIFAR-10 dataset consists of two convolutional layers with 64 $3\times5\times5$ kernels and 64 $64\times5\times5$ kernels and three hidden layers with 1024, 384, 192 neurons, respectively.
\begin{table}[t]
  \centering
  \caption{System Parameters}
  \begin{tabular}{c|c|c|c|c|c|c|c}
    \hline \hline
    \textbf{Parameter} & \textbf{Value} & \textbf{Parameter} & \textbf{Value} & \textbf{Parameter} & \textbf{Value} & \textbf{Parameter} & \textbf{Value} \\ \hline
    $B$ & 1 MHz & $p$ & 0.2 W & $K$ & 4 & $\zeta$ & 1 \\ \hline
    $N_0$ & -174 dBm/Hz & $\alpha$ & $10^{-26}$ & $\gamma^{\rm FEMNIST}$ & 1000 & $\gamma^{\rm CIFAR-10}$ & 2000 \\ \hline
    $f^{\min}$ & $ 2 \times 10^8$ Hz & $f^{\max}$ & $1 \times 10^9$ Hz &
    $\tau$ & 6 & $\tau^{\rm e}$ & 2 \\ \hline
    $T^{\rm FEMNIST, max}$ & 0.02 s & $T^{\rm CIFAR-10, max}$ & 0.05 s &
    $Z^{\rm FEMNIST}$ & 246590 & $Z^{\rm CIFAR-10}$ & 576778 \\ \hline
    \hline
  \end{tabular}
  \label{table: parameter value}
\end{table}
\par
\textbf{Baselines.} There are four baselines to compare the energy consumption and FL performance: (a) the algorithm without quantization that directly uploads the model (see the line labeled with ``No Quantization''), (b) the channel allocation algorithm that optimizes channel allocation and maximizes the quantization level accordingly (``Channel-Allocate''), (c) the algorithm in \cite{client_time_adaptive} that adapts the quantization level to both clients and the training process based on the principle in \cite{client_time_adaptive} without considering wireless constraints (``Principle \cite{client_time_adaptive}''), (d) the algorithm in \cite{Lyapunov_quantization} that optimizes channel allocation and the quantization level under the assumption that all clients have the same sizes of datasets (``Same-Size \cite{Lyapunov_quantization}''). All curves of algorithms are obtained with the average of 5 experiment results, if not mentioned otherwise.
\par
In Section VI-A, different values of $V$ are tested and proper values are chosen for the FEMNIST dataset and the CIFAR-10 dataset. Accuracy curves and energy consumption curves of all algorithms on the FEMNIST datasets are presented in Section VI-B, and curves on the CIFAR-10 dataset are presented in Section VI-C. Finally, the relationships of the quantization level with both the dataset size and the training process are revealed in Section VI-D.
\subsection{Trade-off between Energy Consumption and Performance}
\begin{figure}[t]
  \centering
  \subfigure[$V$ on the FEMNIST dataset ]{\includegraphics[width=0.25\textwidth]{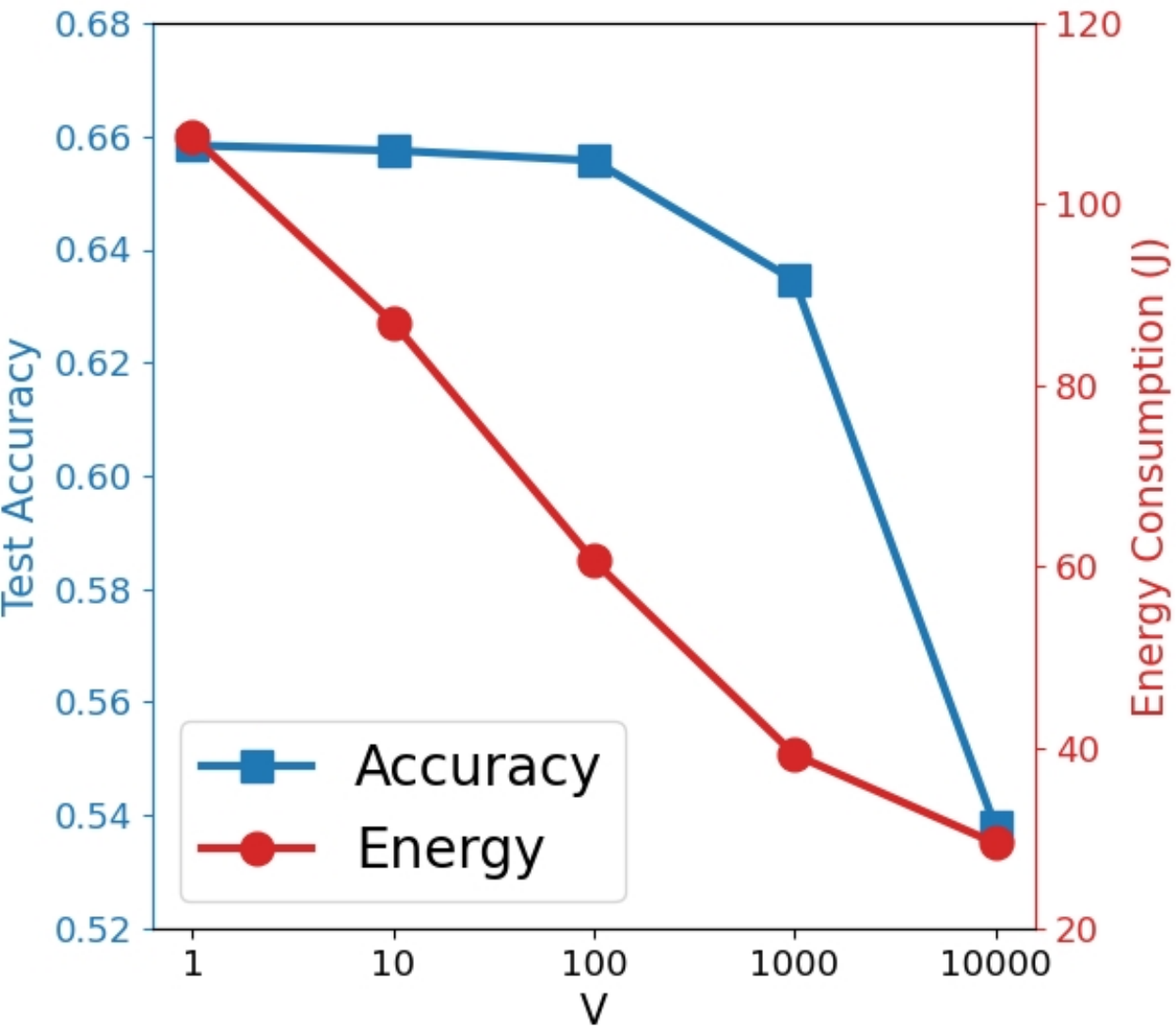}}
  \qquad
  \subfigure[$V$ on the CIFAR-10 dataset
  ]{\includegraphics[width=0.25\textwidth]{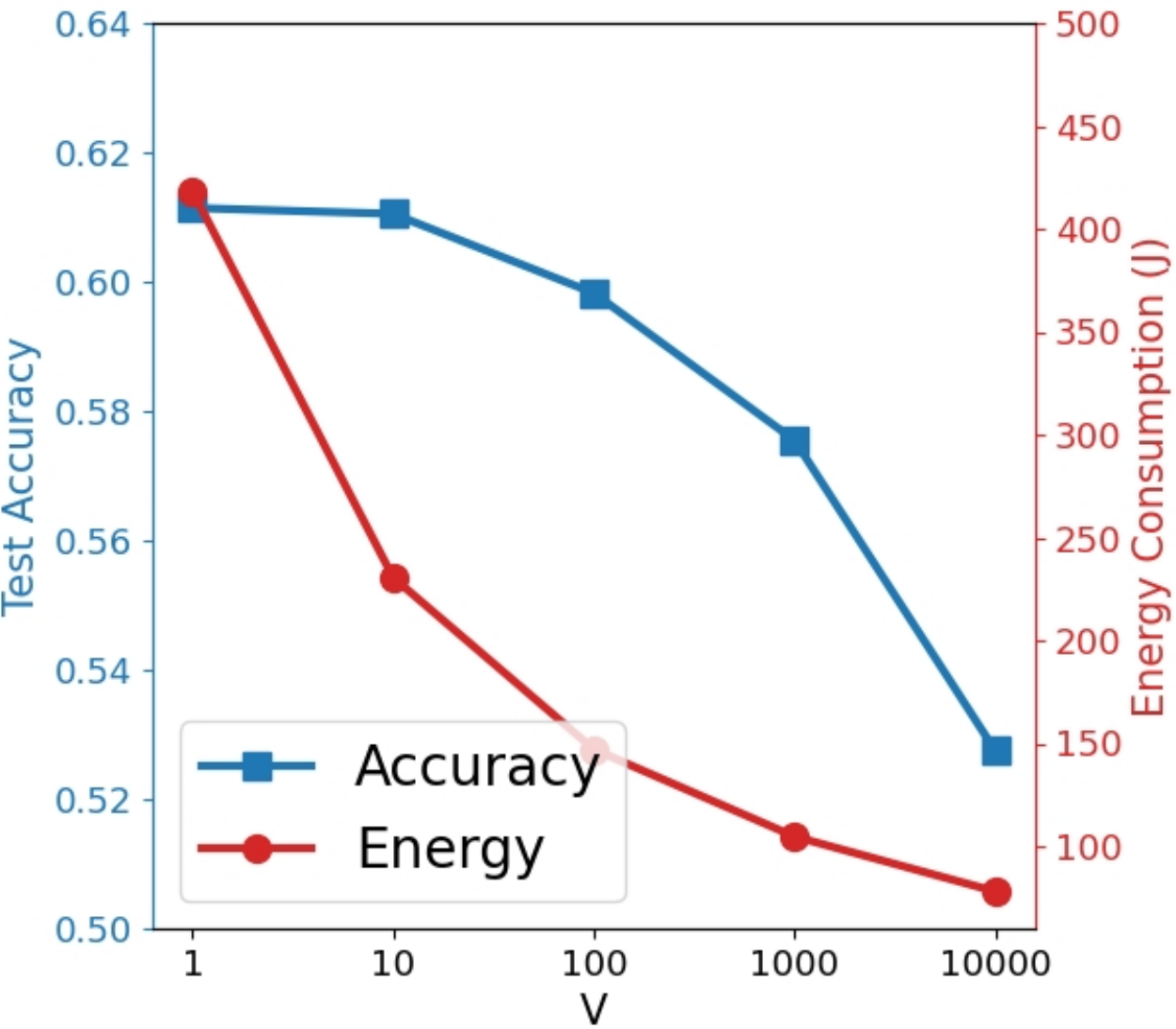}}
  \caption{Test accuracy and accumulated energy consumption curves of the QCCF algorithm with different V values.}
  \label{figure: Trade-off}
\end{figure}
Fig. \ref{figure: Trade-off} plots the impact of $V$ on the accuracy and the energy consumption. It can be seen that both the accuracy and the energy consumption descend as increasing of $V$, which means that a high $V$ saves much energy but acquires terrible FL performance. A low $V$, in turn, pursues excellent FL performance, albeit at the cost of much energy consumption. These conclusions are consistent to our definition in (\ref{equation: Lyapunov drift-plus-penalty}). In a nutshell, the value of $V$ depends on demands of FL performance and energy consumption. The accuracy curve is flat with $V \leq 100$ in Fig. \ref{figure: Trade-off}(a) and $V \leq 10$ in Fig. \ref{figure: Trade-off}(b). Hence, we choose $V=100$ for experiments on the FEMNIST dataset in Section VI-B and $V=10$ for experiments on the CIFAR-10 dataset in Section VI-C.
\subsection{Handwritten Digit and Letter Identification}
\begin{figure}[t]
  \centering
  \subfigure[Accuracy with $\beta=150$ ]{\includegraphics[width=0.23\textwidth]{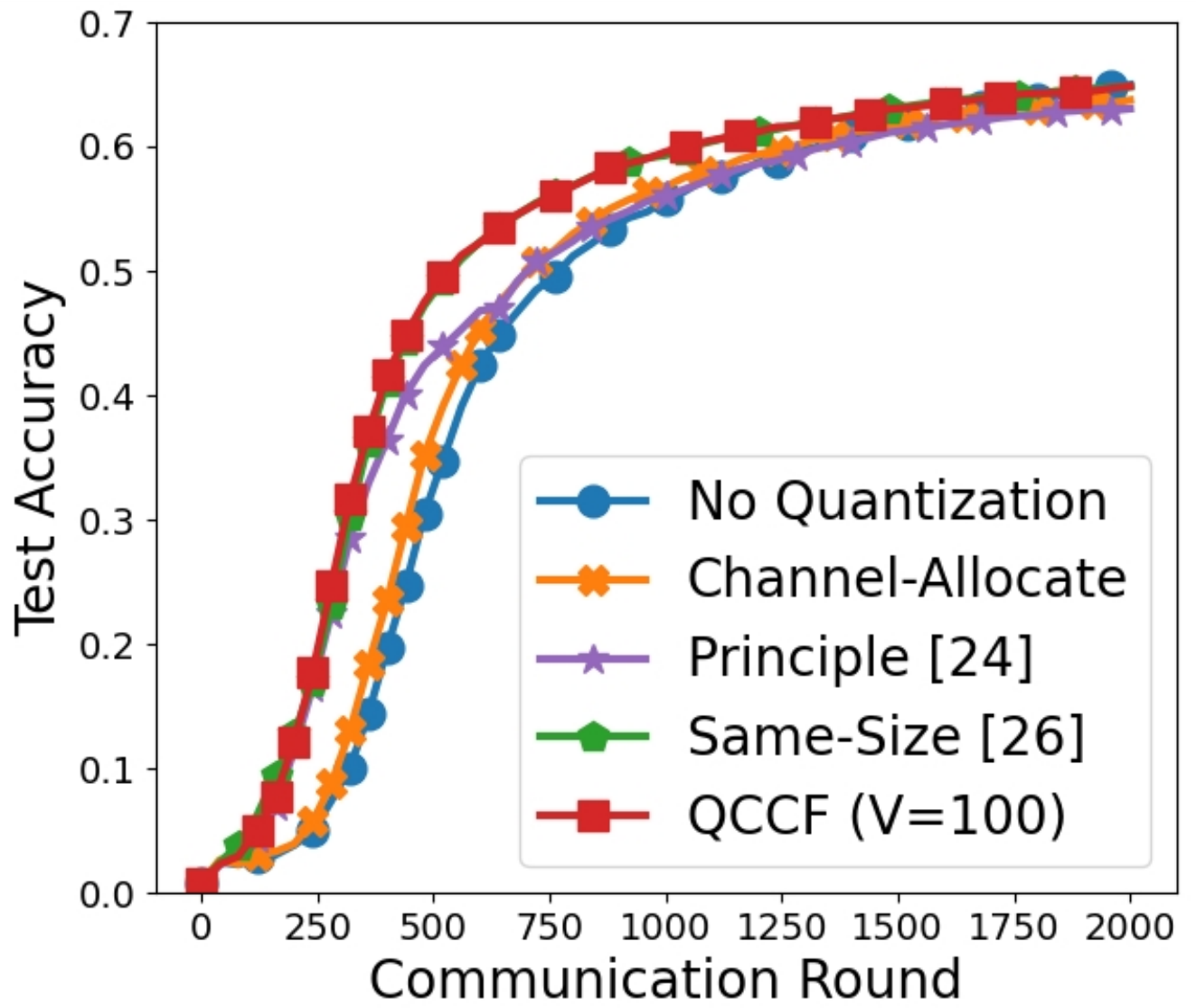}}
  \subfigure[Energy with $\beta=150$
  ]{\includegraphics[width=0.23\textwidth]{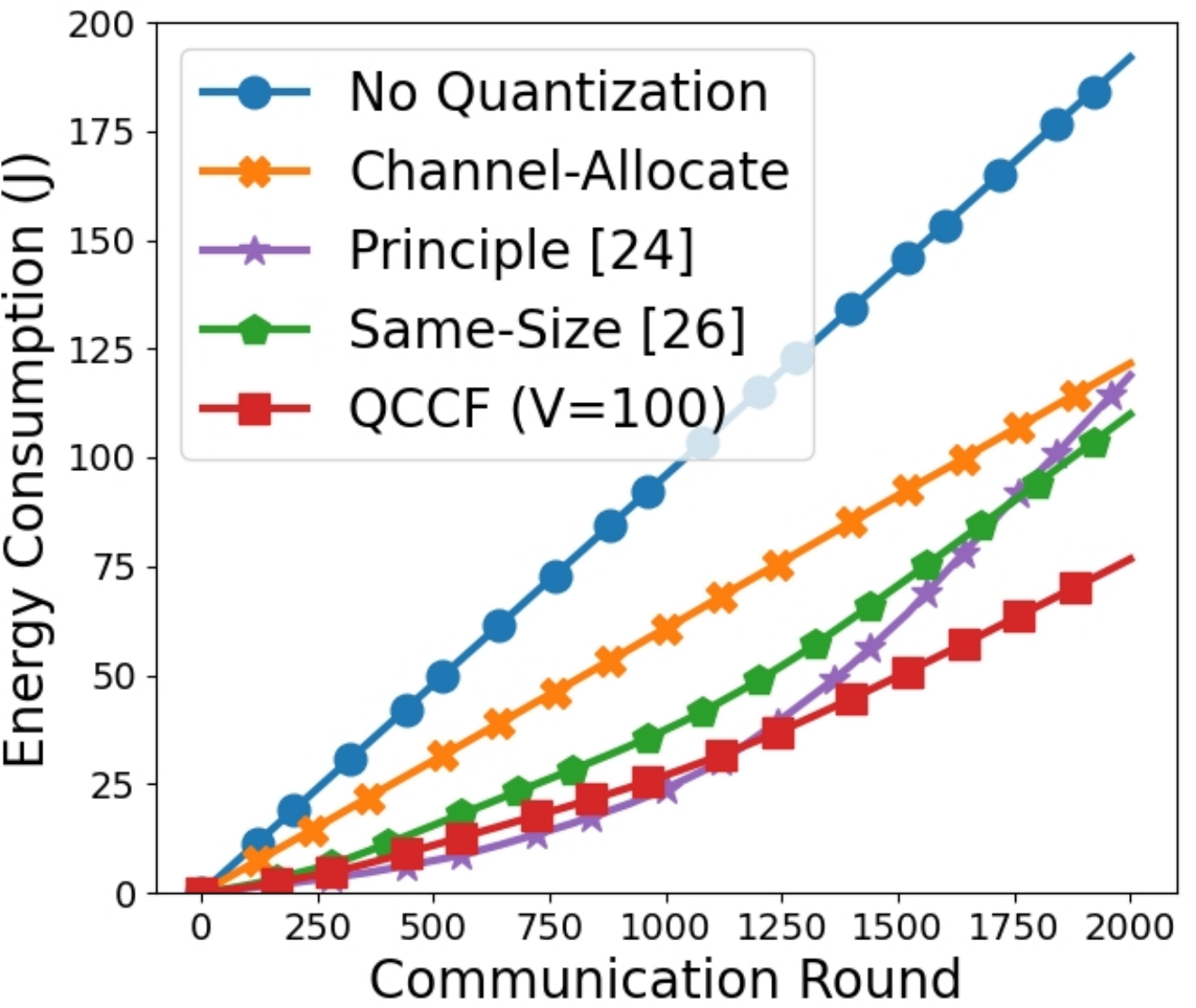}}
  \subfigure[Accuracy with $\beta=300$ ]{\includegraphics[width=0.23\textwidth]{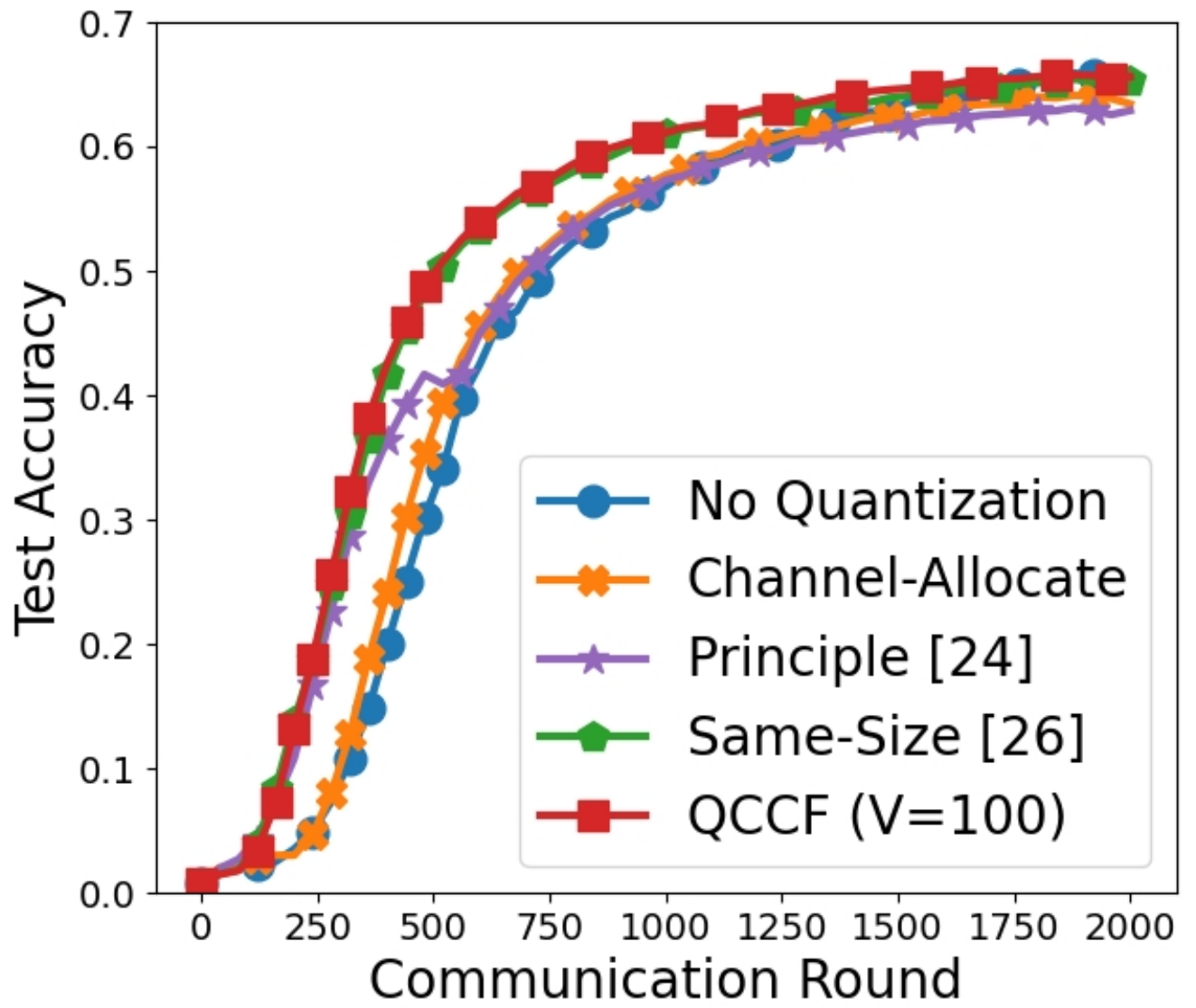}}
  \subfigure[Energy with $\beta=300$
  ]{\includegraphics[width=0.23\textwidth]{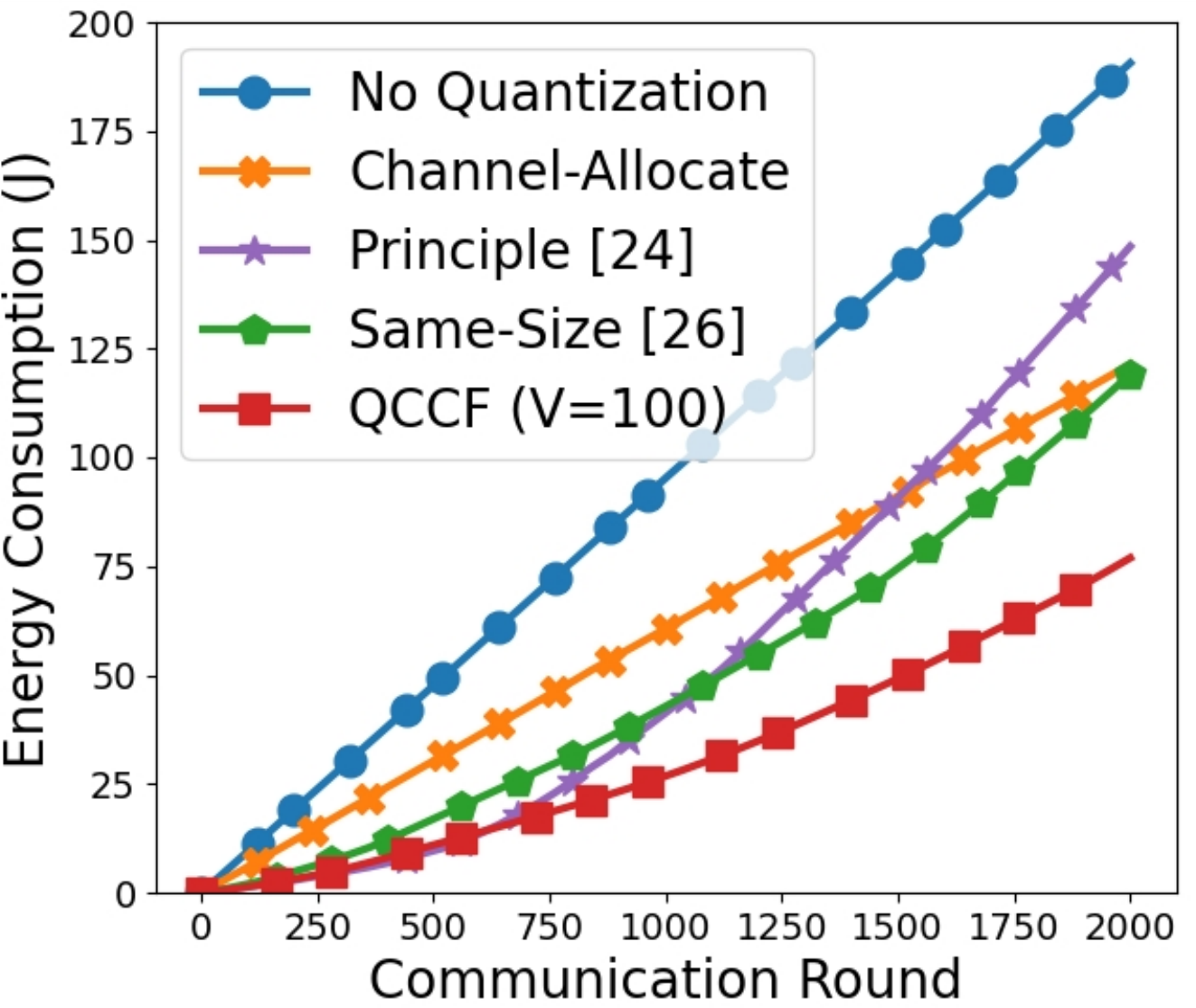}}
  \caption{Test accuracy curves and accumulated energy consumption curves of related algorithms on the FEMNIST dataset.}
  \label{figure: FEMNIST}
\end{figure}

Fig. \ref{figure: FEMNIST} compares FL performance and the energy consumption of our QCCF algorithm with the other four baselines under different standard deviations of sizes on the FEMNIST dataset. From Fig. \ref{figure: FEMNIST}(a) and (c), we can observe that our QCCF algorithm achieves the fastest convergence among all algorithms. And in Fig. \ref{figure: FEMNIST}(b) and (d), it can be noticed that our QCCF algorithm consumes less energy than other baselines. Comparing Fig. \ref{figure: FEMNIST}(b) with (d), we can see that the principle algorithm and the same-size algorithm with $\beta = 300$ consume more energy than $\beta = 150$. This is because the principle algorithm sets high quantization levels for clients with large datasets, resulting in much energy consumption to accomplish computation and communication within latency. As for the same-size algorithm, computation latency is determined by the largest dataset under the same-size assumption. Hence, all clients accelerate CPUs to satisfy the latency constraint, which increases energy consumption as $\beta$ rises. Our QCCF algorithm, by contrast, reduces quantization levels and CPU frequencies of clients with large datasets, so that stable energy consumption is kept with $\beta = 150, 300$. From Fig. \ref{figure: FEMNIST}(a) and (c), we can observe that the increase of the principle algorithm abnormally gets too slow in the later part of the training process. This is due to the fact that the quantization level keeps rising according to the principle, and the quantization levels of clients with large datasets are too high to participate within latency. These clients dropping out leads to inadequate training of the global model on their datasets, which accounts for the slow increase of the principle algorithm in the later part.
\subsection{Colored Image Identification}
\begin{figure}[t]
  \centering
  \subfigure[Accuracy with $\beta=150$ ]{\includegraphics[width=0.23\textwidth]{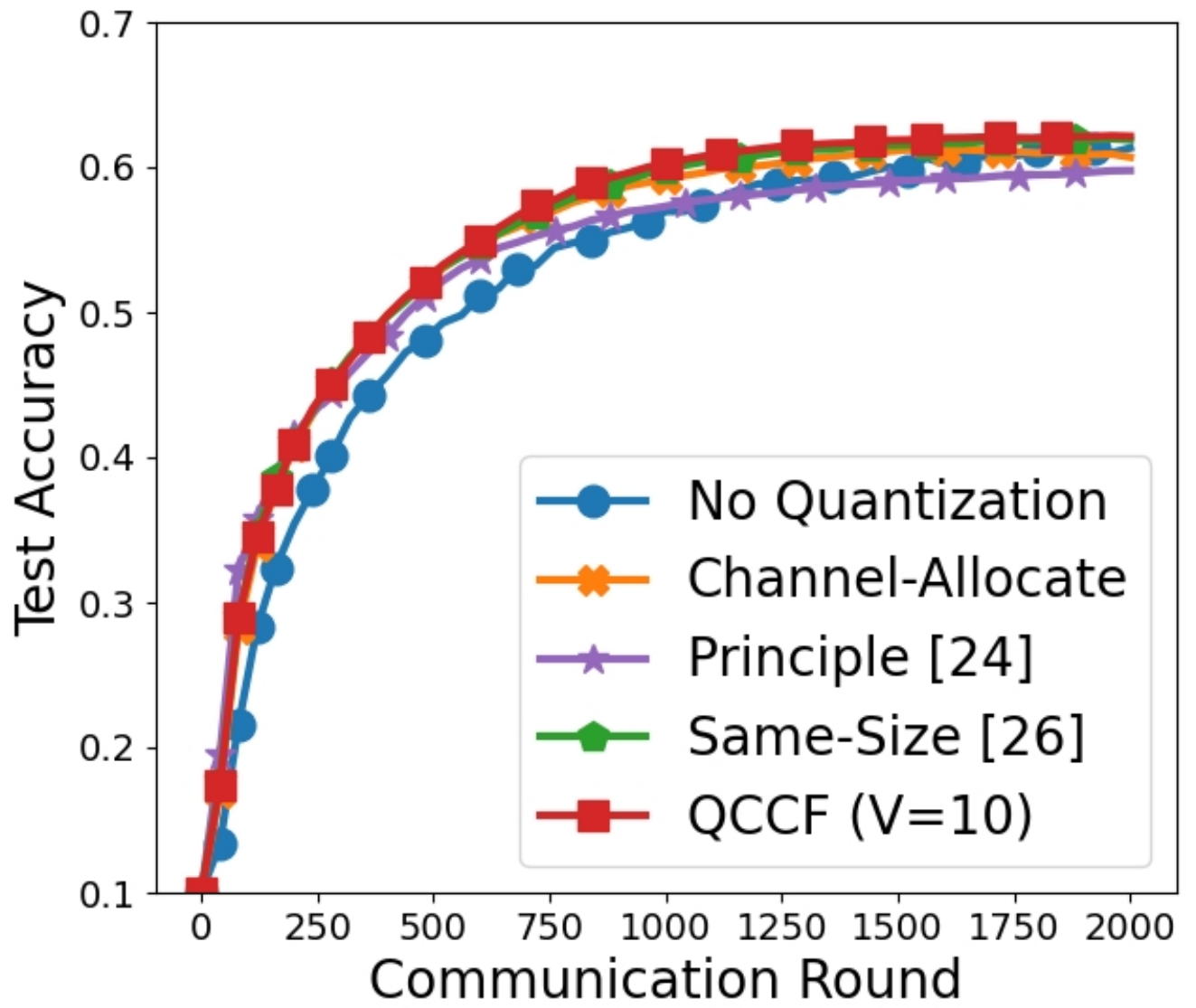}}
  \subfigure[Energy with $\beta=150$
  ]{\includegraphics[width=0.23\textwidth]{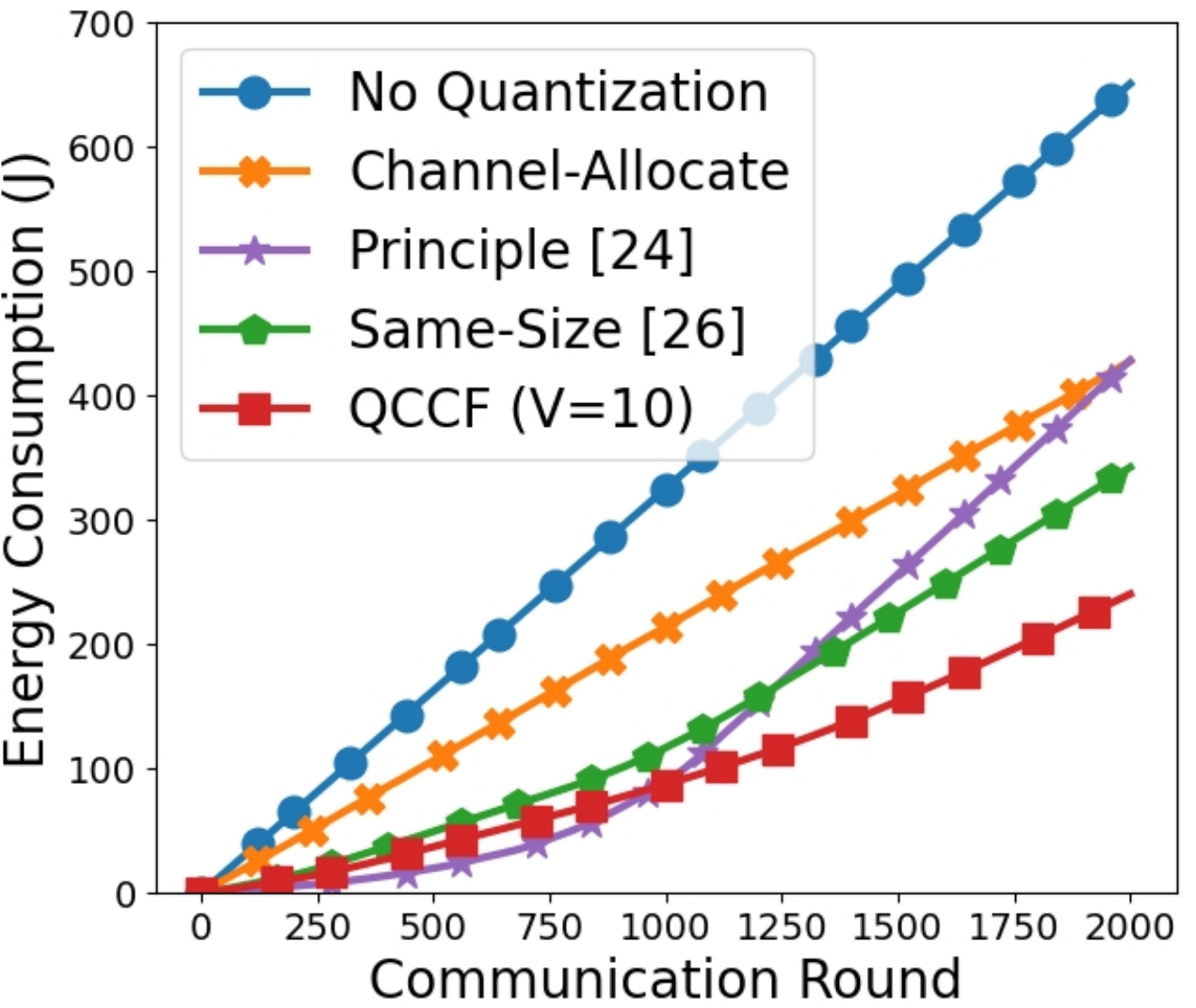}}
  \subfigure[Accuracy with $\beta=300$ ]{\includegraphics[width=0.23\textwidth]{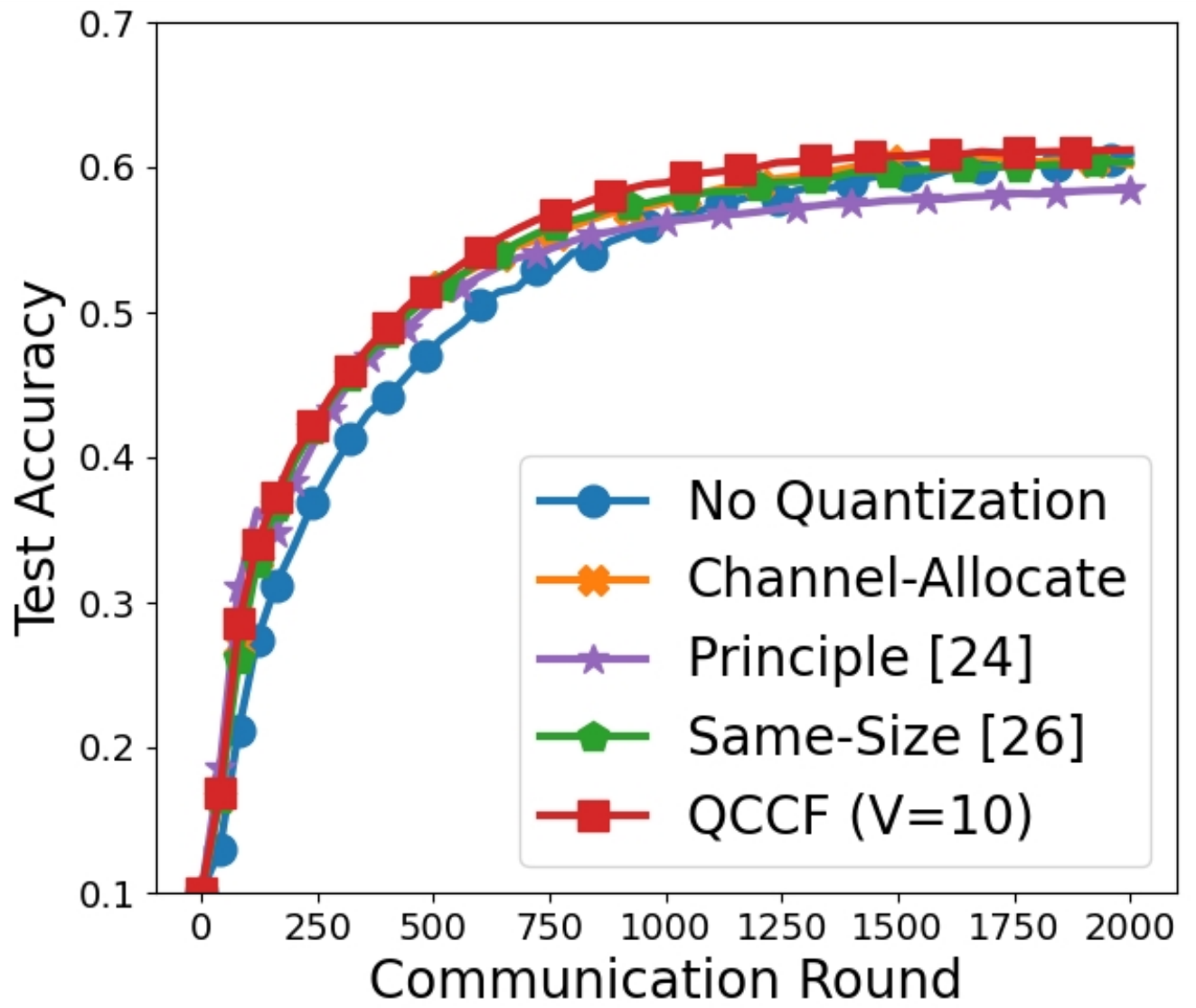}}
  \subfigure[Energy with $\beta=300$
  ]{\includegraphics[width=0.23\textwidth]{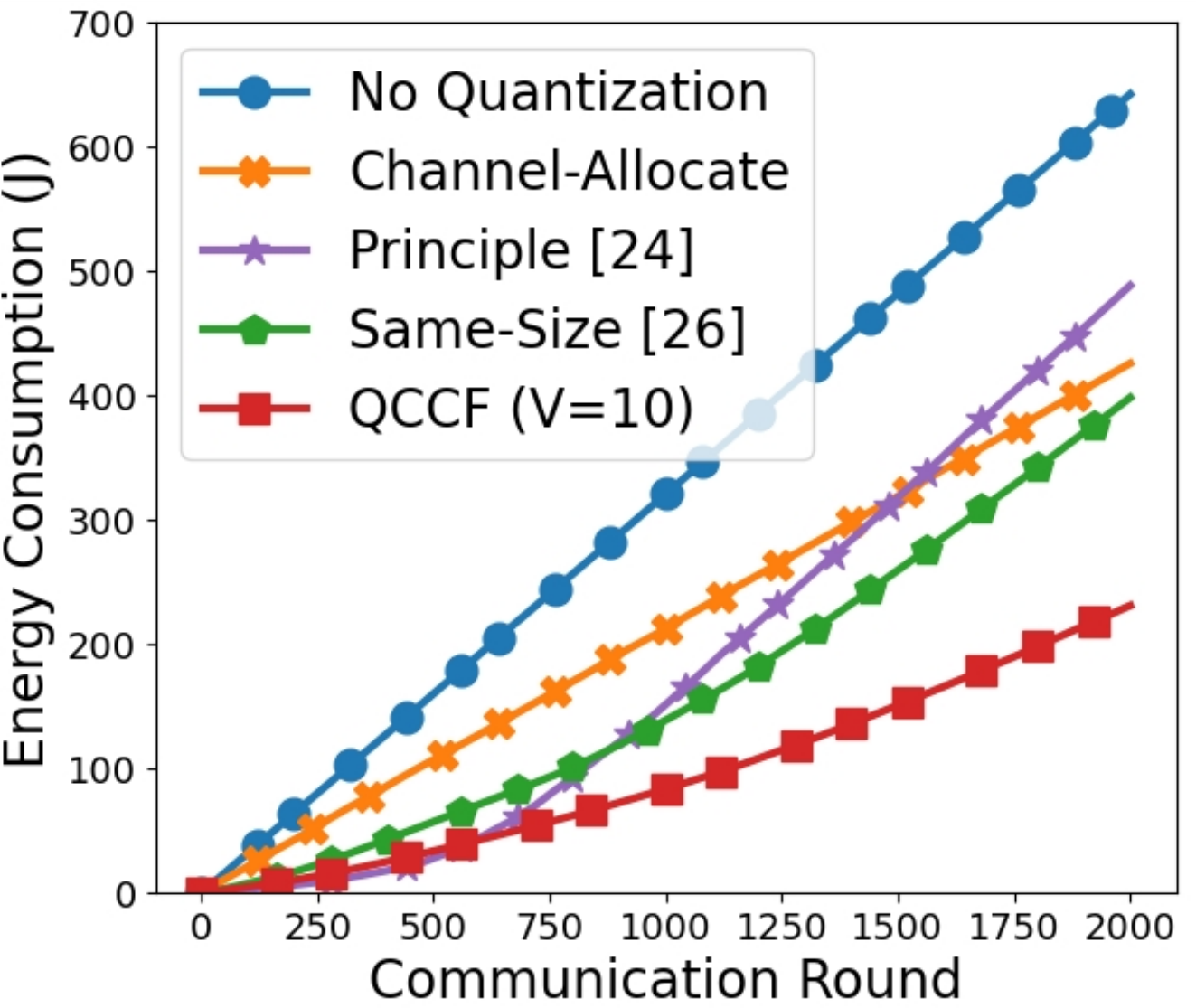}}
  \caption{Test accuracy curves and accumulated energy consumption curves of related algorithms on the CIFAR-10 dataset.}
  \label{figure: CIFAR10}
\end{figure}

Fig. \ref{figure: CIFAR10} depicts FL performance and energy consumption of the five algorithms under different standard deviations of sizes on the CIFAR-10 dataset. In Fig. \ref{figure: CIFAR10}, it is obvious that our QCCF algorithm converges fastest and consumes least the energy among all algorithms. And Fig. \ref{figure: CIFAR10}(d) reveals that the principle algorithm and the same-size algorithm can not adapt to the datasets with large $\beta$. These conclusions are similar to those in Section VI-B. Fig. \ref{figure: CIFAR10}(a) and (c) both show that the principle algorithm increases slowly in the later part of the training process. Its reason is the same in Section VI-B, and the complex task on the CIFAR-10 dataset aggravates its performance, suggesting a proper design of the quantization level in wireless FL considers not only theoretical FL performance but also real wireless constraints.
\subsection{Analysis of Quantization Level}
\begin{figure}[t]
  \centering
  \subfigure[Quantization levels varying \textcolor{white}{a} with communication rounds ]{\includegraphics[height=0.15\textheight]{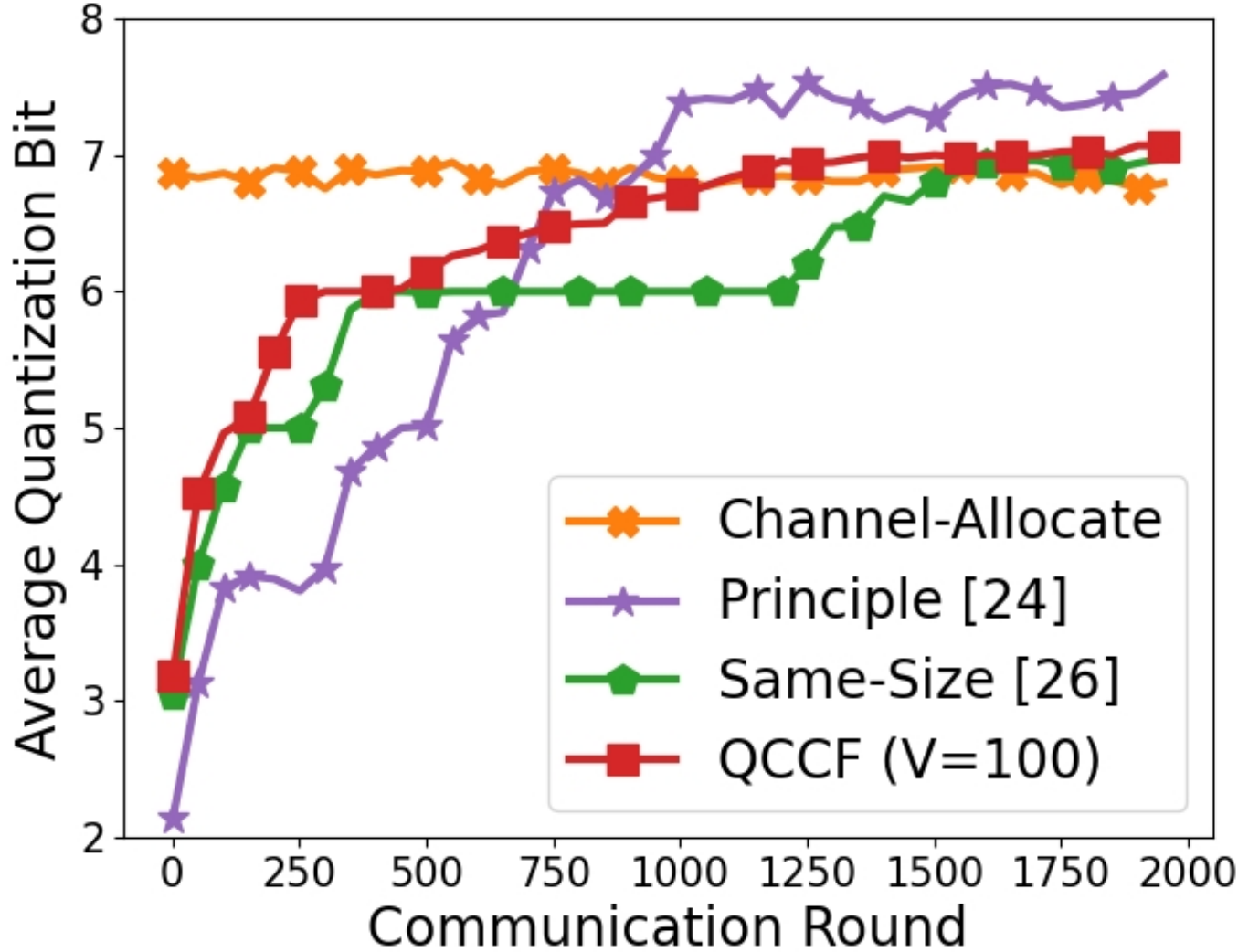}}
  \qquad
  \subfigure[Quantization levels varying among clients with different dataset sizes ]{\includegraphics[height=0.15\textheight]{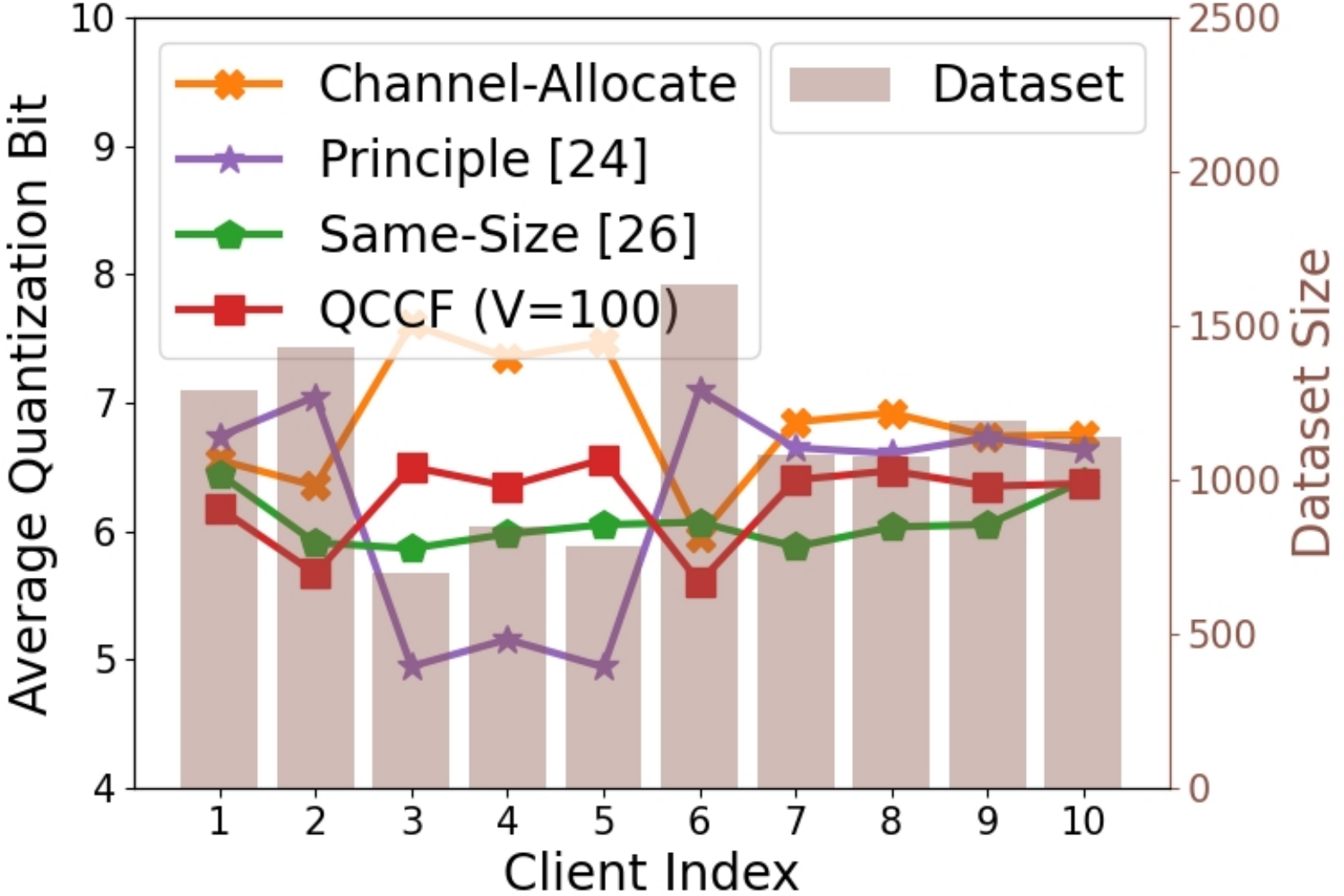}}
  \caption{The relationships of quantization levels with the training process and dataset sizes for related algorithms.}
  \label{figure: Quantization}
\end{figure}

Fig. \ref{figure: Quantization} plots varying quantization levels of different algorithms with respect to the training process and clients. Due to the dataset size vector can not be the same in different experiments, the result of an experiment on the FEMNIST dataset is provided. Despite our analysis based on one experiment, we also find the following conclusions are suitable for other experiments.
\par
Fig. \ref{figure: Quantization}(a) shows the relationship between the quantization level and the training process. It can be noticed that the principle algorithm, the same-size algorithm, and our QCCF algorithm all increase the quantization level with the training process. And the final quantization level of the principle algorithm is larger than the other two. The underlying reason is that the principle algorithm blindly keeps the quantization level rising and leads to clients dropping out, which not only consumes more energy but also achieves poor performance. As for the channel-allocate algorithm, there are no significant differences in quantization levels in all communication rounds. This result is rather intuitive since the quantization level of the channel-allocate algorithm only relies on channel states, which follow the same distribution in different communication rounds.
\par
In Fig. \ref{figure: Quantization}(b), we show the relationship between the quantization level and dataset sizes. From this figure, we can see that the quantization levels of the channel-allocate algorithm and our QCCF algorithm are negatively related to dataset sizes. It attributes to the fact that these two algorithms consider both the latency constraint and different dataset sizes among clients. The same-size algorithm ignores the latter so that its quantization level is unrelated to dataset sizes. And according to the quantization principle of the principle algorithm, the quantization level is proportional to the dataset size, which can be validated by Fig. \ref{figure: Quantization}(b).
\par
After the above analysis, we conclude that low energy consumption and good FL performance mainly benefit from the quantization level increasing with the training process and being negatively related to dataset sizes, i.e., \textbf{\emph{Remark 1}} and \textbf{\emph{Remark 2}} in Section V-C. The rising quantization level caters to the need for model convergence and substantially reduces communication energy consumption in the early training process. The negative relationship compensates divergent computation latency to guarantee the participation of FL and even allows clients with large datasets not to accelerate CPUs, which also saves computation energy indirectly.
\section{Conclusion}
This paper has proposed a novel approach to reduce energy consumption in wireless FL with the adaption of the quantization level to both the training process and different dataset sizes of clients. First, we have developed an FL framework with client scheduling and the doubly adaptive quantization level. To ensure the convergence of our FL framework, we have derived an upper bound with the quantization error part and the data property part. These two parts and other wireless settings have served as constraints. And a long-term optimization problem has been formulated to minimize total energy consumption by designing quantization levels, scheduling clients, allocating channels, and controlling CPU frequencies. Through Lyapunov optimization, the long-term problem has been transformed into an instantaneous problem, which can be decomposed into two subproblems and solved by KKT conditions and the genetic algorithm. The closed-form solution has suggested that the doubly adaptive quantization level should rise with the training process and be negatively correlated with dataset sizes, which has been validated in later experiments. Furthermore, experiment results have demonstrated that our QCCF algorithm outperforms other listed algorithms in terms of both energy consumption and FL performance.
\par
There are some directions for our future works, for instance, the doubly adaptive quantization for updates. And other quantization methods such as the block floating point quantization and the sequence mapping quantization, may also be integrated with the doubly adaptive quantization.
\appendix
\subsection{Proof of Lemma \ref{lemma: differences}\label{proof1}}
With local gradients, the sum of differences between local models and the initial model can be added an zero term and expanded into
\begin{align}
& \sum_{i = 1}^U w_i^n \| \bm \theta_i^{n, m} - \bm \psi^{n, 0} \|^2 = \eta^2 \sum_{i=1}^U w_i^n \Big\| \sum_{t=0}^{m-1} \nabla F_i(\bm \theta_i^{n,t}, \xi_i^{n,t}) - \sum_{t=0}^{m-1} \nabla F_i(\bm \theta_i^{n,t}) + \sum_{t=0}^{m-1} \nabla F_i(\bm \theta_i^{n,t}) \Big\|^2 \notag \\
& = \eta^2 \sum_{i=1}^U w_i^n \Big\| \sum_{t=0}^{m-1} \nabla F_i(\bm \theta_i^{n,t}, \xi_i^{n,t}) - \sum_{t=0}^{m-1} \nabla F_i(\bm \theta_i^{n,t}) \Big\|^2 + \eta^2 \sum_{i=1}^U w_i^n \Big\| \sum_{t=0}^{m-1} \nabla F_i(\bm \theta_i^{n,t}) \Big\|^2 \label{equation: lemma2_1} \\
& \quad + 2\eta^2 \sum_{i=1}^U w_i^n \Big\langle \sum_{t=0}^{m-1} \nabla F_i(\bm \theta_i^{n,t}, \xi_i^{n,t}) - \sum_{t=0}^{m-1} \nabla F_i(\bm \theta_i^{n,t}), \sum_{t=0}^{m-1} \nabla F_i(\bm \theta_i^{n,t}) \Big\rangle. \notag
\end{align}
With the independence among $\xi_i^{n, 0}, \xi_i^{n,1}, \cdots, \xi_i^{n, m-1}$, we take the expectation of mini-batches as
\begin{equation}
\sum_{i = 1}^U w_i^n \mathbb E \left[\| \bm \theta_i^{n, m} - \bm \psi^{n, 0} \|^2 \right] \overset{(a)}{\leq} \eta^2 \sum_{i=1}^U w_i^n \sum_{t=0}^{m-1} (\sigma_i^n)^2+ \eta^2 \sum_{i=1}^U w_i^n \Big\| \sum_{t=0}^{m-1} \nabla F_i(\bm \theta_i^{n,t}) \Big\|^2,
\label{equation: lemma2_2}
\end{equation}
where $(a)$ comes from \textbf{Assumption \ref{assumption: mini-batch}} and variance additivity of independent random variables.
Next, based on $\| \bm x + \bm y \|^2 \leq 2\|\bm x\|^2 + 2\|\bm y\|^2$ and Jensen's inequality, the norm of accumulative gradients in the last line of (\ref{equation: lemma2_2}) is transformed into
\begin{equation}
\begin{aligned}
& \sum_{i=1}^U w_i^n \left\| \sum_{t=0}^{m-1} \nabla F_i(\bm \theta_i^{n,t}) \right\|^2 = \sum_{i=1}^U w_i^n \left\| \sum_{t=0}^{m-1} \left(\nabla F_i(\bm \theta_i^{n,t}) - \nabla F_i(\bm \psi^{n,t}) + \nabla F_i(\bm \psi^{n,t})\right) \right\|^2 \\
& \leq 2m \sum_{i=1}^U w_i^n \sum_{t=0}^{m-1} \left\| \nabla F_i(\bm \theta_i^{n,t}) - \nabla F_i(\bm \psi^{n,t}) \right\|^2 + 2m \sum_{i=1}^U w_i^n \sum_{t=0}^{m-1} \left\| \nabla F_i(\bm \psi^{n,t}) \right\|^2.
\end{aligned}
\label{equation: lemma2_3}
\end{equation}
To enlarge the difference of gradients, we can derive $ \sum^U_{i=1} w_i^n \|\bm \theta_i^{n,t} - \bm \psi^{n,m}\|^2 \leq \sum^U_{i=1} w_i^n \|\bm \theta_i^{n,t} - \bm \psi^{n,0}\|^2 $ according to the \textbf{Definition \ref{defintion}}. And more updates mean larger differences from the initial model, that is, $\| \bm \theta_i^{n,t} - \bm \psi^{n,0} \| \leq \| \bm \theta_i^{n,m} - \bm \psi^{n,0} \|$ for $t \leq m$. With the previous inequalities and \textbf{Assumption \ref{assumption: gradient}}, hence, the norm of accumulative gradients is bounded by
\begin{equation}
\sum_{i=1}^U w_i^n \Big\| \sum_{t=0}^{m-1} \nabla F_i(\bm \theta_i^{n,t}) \Big\|^2 \leq 2m^2L^2 \sum_{i=1}^U w_i^n \left\| \bm \theta_i^{n,m} - \bm \psi^{n,0} \right\|^2 + 2m^2 \sum_{i=1}^U w_i^n \left(G_i^n\right)^2.
\label{equation: lemma2_4}
\end{equation}\par
Substituting (\ref{equation: lemma2_4}) into (\ref{equation: lemma2_2}), we have
$
\sum_{i = 1}^U w_i^n \mathbb E \left[\| \bm \theta_i^{n, m} - \bm \psi^{n, 0} \|^2 \right] \leq \eta^2 \sum_{i=1}^U w_i^n \sum_{t=0}^{m-1} (\sigma_i^n)^2 + 2m^2\eta^2\sum_{i=1}^U w_i^n \left(G_i^n\right)^2 + 2m^2\eta^2 L^2 \sum_{i=1}^U w_i^n \mathbb E [\left\| \bm \theta_i^{n,m} - \bm \psi^{n,0} \right\|^2].$
Then, rearranging terms and dividing both sides by $(1-2m^2\eta^2L^2)$, we obtain (\ref{equation: upper bound model}). This completes the proof of \textbf{Lemma \ref{lemma: differences}}.
\subsection{Proof of Theorem \ref{theorem: update} \label{proof2}}
We consider the variation of the loss function after a update. Hence, according to \textbf{Assumption \ref{assumption: smooth}}, the difference of loss functions are expanded into
\begin{equation}
F(\bm \psi^{n, m+1}) - F(\bm \psi^{n,m}) \leq \big\langle \nabla F(\bm \psi^{n,m}), \bm \psi^{n,m+1} - \bm \psi^{n,m} \big\rangle + \frac{L}{2}\|\bm \psi^{n,m+1} - \bm \psi^{n,m}\|^2.
\label{equation: update L-smooth}
\end{equation}
The key to derive the upper bound of the gradient is to transform the above two terms. Thus, the cross term is firstly taken into consideration. Taking the expectation of all independent mini-batches $\xi_1^{n,m}, \xi_2^{n,m}, \cdots, \xi_U^{n,m}$, we have
\begin{align}
& \mathbb E \left[\big\langle \nabla F(\bm \psi^{n,m}), \bm \psi^{n,m+1} - \bm \psi^{n,m} \big\rangle\right] \overset{(a)}{=} -\eta \big\langle \nabla F(\bm \psi^{n,m}),  \sum^U_{i=1}w_i^n \nabla F_i(\bm \theta_i^{n,m}) \big\rangle \label{equation: cross term expectation} \\
& \overset{(b)}{=} \frac{\eta}{2} \Big\| \nabla F(\bm \psi^{n,m}) - \sum^U_{i=1}w_i^n \nabla F(\bm \theta^{n,m}_i) \Big\|^2 -\frac{\eta}{2} \left\| \nabla F(\bm \psi^{n,m}) \right\|^2 - \frac{\eta}{2} \Big\| \sum^U_{i=1}w_i^n \nabla F_i(\bm \theta_i^{n,m}) \Big\|^2, \notag
\end{align}
where $(a)$ is due to \textbf{Assumption \ref{assumption: mini-batch}}, and $(b)$ results from $- \langle\bm x, \bm y\rangle = \frac{\|\bm x - \bm y \|^2 - \|\bm x\|^2 - \|\bm y\|^2}{2}$. Then we consider the first term in $(b)$ of (\ref{equation: cross term expectation}). Adding a zero term and utilizing $\| \bm x - \bm y \|^2 \leq 2\|\bm x\|^2 + 2\|\bm y\|^2$, we transform the formula into
$
\frac{\eta}{2} \Big\| \nabla F(\bm \psi^{n,m}) - \sum^U_{i=1}w_i^n \nabla F(\bm \theta^{n,m}_i) \Big\|^2 \leq \eta \Big\| \sum^U_{i=1}  (w_i - w_i^n) \nabla F_i(\bm \psi^{n,m}) \Big\|^2 + \eta \Big\| \sum^U_{i=1}  w_i^n \left(\nabla F_i(\bm \psi^{n,m}) - \nabla F_i(\bm \theta_i^{n,m})\right) \Big\|^2.
$ It is noted that the aggregation weight serves as a scalar about clients, which makes norms fairly complicated. Hence, aggregation weights are extracted and we have
\begin{equation}
\begin{aligned}
& \frac{\eta}{2} \Big\| \nabla F(\bm \psi^{n,m}) - \sum^U_{i=1}w_i^n \nabla F(\bm \theta^{n,m}_i) \Big\|^2 \\
& \overset{(a)}{\leq} 4\eta \left( \sum^U_{i=1} \frac{|w_i - w_i^n|}{2 - 2 \sum_{j \in \mathcal U^n_{\rm in}} w_j} \left\|\nabla F_i(\bm \psi^{n,m})\right\| \right)^2 + \eta \left\| \sum^U_{i=1}  w_i^n \left(\nabla F_i(\bm \psi^{n,m}) - \nabla F_i(\bm \theta_i^{n,m})\right) \right\|^2 \\
& \overset{(b)}{\leq} 4 \eta \sum^U_{i=1} \frac{|w_i - w_i^n|}{2 - 2 \sum_{j \in \mathcal U^n_{\rm in}} w_j} \left\|\nabla F_i(\bm \psi^{n,m})\right\|^2 + \eta L^2 \sum^U_{i=1} w_i^n \|\bm\psi^{n,m} - \bm\theta_i^{n,m}\|^2 \\
& \overset{(c)}{\leq} 2 \eta \sum^U_{i=1} (1-a_i^nw_i)(G_i^n)^2 + \eta L^2 \sum^U_{i=1} w_i^n \|\bm\psi^{n,m} - \bm\theta_i^{n,m}\|^2,
\end{aligned}
\label{equation: out weight}
\end{equation}
where $(a)$ is according to $\sum^U_{j = 1} |w_j - w_j^n| = \sum_{j \in \mathcal U_{\rm in}^n} (w_j^n - w_j) + \sum_{j \in \mathcal U_{\rm out}^n} w_j  = 2 - 2\sum_{j \in \mathcal U^n_{\rm in}} w_j \leq 2$ and adding a zero term $\sum^U_{i=1} w_i^n \nabla F_i(\bm \psi^{n,m}) - \sum^U_{i=1} w_i^n \nabla F_i(\bm \psi^{n,m})$, $(b)$
is due to Jensen's inequality and \textbf{Assumption \ref{assumption: smooth}}, and $(c)$ is based on $|w_i - w_i^n| \leq (1 - a_i^nw_i)$ for $a_i^n = 0, 1$. \par
Now we can substitute (\ref{equation: out weight}) into (\ref{equation: cross term expectation}) and we have
\begin{equation}
\begin{aligned}
& \mathbb E \left[\big\langle \nabla F(\bm \psi^{n,m}), \bm \psi^{n,m+1} - \bm \psi^{n,m} \big\rangle\right] \leq 2 \eta \sum^U_{i=1} (1-a_i^nw_i) (G_i^n)^2 \\
& \qquad + \eta L^2 \sum^U_{i=1} w_i^n \|\bm\psi^{n,m} - \bm\theta_i^{n,m}\|^2  -\frac{\eta}{2} \left\| \nabla F(\bm \psi^{n,m}) \right\|^2 - \frac{\eta}{2} \Big\| \sum^U_{i=1}w_i^n \nabla F_i(\bm \theta_i^{n,m}) \Big\|^2.
\end{aligned}
\label{equation: cross term substitution}
\end{equation}
In (\ref{equation: cross term substitution}), $\sum^U_{i=1} w_i^n \|\bm\psi^{n,m} - \bm\theta_i^{n,m}\|^2$ needs the initial model parameter to construct the form in \textbf{Lemma \ref{lemma: differences}}. Hence, $\bm \psi ^{n,0} - \bm \psi ^{n,0}$ is added and we have
\begin{align}
& \sum^U_{i=1} w_i^n \|\bm\psi^{n,m} - \bm\theta_i^{n,m}\|^2 \label{equation: model difference} = \sum^U_{i=1} w_i^n \|(\bm\psi^{n,m} - \bm\psi^{n,0}) + (\bm\psi^{n,0} - \bm\theta_i^{n,m})\|^2\\
& \overset{(a)}{=} \sum^U_{i=1} w_i^n \| \bm\theta_i^{n,m} - \bm\psi^{n,0} \|^2 - \sum^U_{i=1} w_i^n \|\bm\psi^{n,m} - \bm\psi^{n,0}\|^2 \overset{(b)}{\leq} \sum^U_{i=1} w_i^n \|\bm\theta_i^{n,m} - \bm\psi^{n,0}\|^2, \notag
\end{align}
where $(a)$ follows \textbf{Definition \ref{defintion}} and $(b)$ is because that the second term in $(a)$ is negative so that it can be ignored. Now, it is time to utilize \textbf{Lemma \ref{lemma: differences}} and substitute (\ref{equation: model difference}) into (\ref{equation: cross term substitution}) to obtain
\begin{equation}
\begin{aligned}
& \mathbb E \left[\big\langle \nabla F(\bm \psi^{n,m}), \bm \psi^{n,m+1} - \bm \psi^{n,m} \big\rangle\right] \leq 2 \eta \sum^U_{i=1} (1-a_i^nw_i) (G_i^n)^2 + \\
& \qquad \eta L^2 \sum^U_{i=1} w_i^n \|\bm\theta_i^{n,m} - \bm\psi^{n,0}\|^2  -\frac{\eta}{2} \left\| \nabla F(\bm \psi^{n,m}) \right\|^2 - \frac{\eta}{2} \Big\| \sum^U_{i=1}w_i^n \nabla F_i(\bm \theta_i^{n,m}) \Big\|^2.
\end{aligned}
\label{equation: cross term final}
\end{equation}\par
So far, the derivation of the cross term in (\ref{equation: update L-smooth}) has been finished. Next, the model difference in (\ref{equation: update L-smooth}) will be analyzed. With local gradients, the model difference is expanded into
$
\|\bm\psi^{n,m+1} - \bm\psi^{n,m}\|^2 = \eta^2 \big\| \sum^U_{i = 1} w_i^n \left(\nabla F_i(\bm\theta_i^{n,m}, \xi_i^{n,m}) - \nabla F_i(\bm\theta_i^{n,m})\right) + \sum^U_{i = 1} w_i^n \nabla F_i(\bm\theta_i^{n,m})\big\|^2.
$
Futhermore, based on \textbf{Assumption \ref{assumption: mini-batch}} and independence between $\xi_1^{n,m}, \xi_2^{n,m}, \cdots, \xi_U^{n,m}$, the expectation of mini-batches is given by
\begin{equation}
\mathbb E\left[\|\bm\psi^{n,m+1} - \bm\psi^{n,m}\|^2\right] = \frac{\eta^2 L}{2} \sum^U_{i=1} w_i^n (\sigma_i^n)^2 + \frac{\eta^2 L}{2} \Big\| \sum^U_{i = 1} w_i^n \nabla F_i(\bm\theta_i^{n,m}) \Big\|^2. \label{equation: global model difference expectation}
\end{equation}
\par
In the end, (\ref{equation: cross term substitution}) and (\ref{equation: global model difference expectation}) are substituted into (\ref{equation: update L-smooth}) as
\begin{equation}
\begin{aligned}
& \mathbb E[F(\bm \psi^{n, m+1}) - F(\bm \psi^{n,m})] \leq 2 \eta \sum^U_{i=1} (1-a_i^nw_i) (G_i^n)^2 + \eta L^2 \sum^U_{i=1} w_i^n \|\bm\psi^{n,m} - \bm\theta_i^{n,m}\|^2  \\
& \qquad + \frac{\eta^2 L -\eta}{2} \left\| \sum^U_{i = 1} w_i^n \nabla F_i(\bm\theta_i^{n,m}) \right\|^2 + \frac{\eta^2 L}{2} \sum^U_{i=1} w_i^n (\sigma_i^n)^2 -\frac{\eta}{2} \mathbb E\left[\left\| \nabla F(\bm \psi^{n,m}) \right\|^2\right].
\end{aligned}
\label{equation: update substitution}
\end{equation}
It is noticed that the sign of $\frac{\eta^2 L -\eta}{2}$ is negative with $\eta L <1$. Hence, this term can be ignored, and (\ref{equation: update substitution}) is rearranged into (\ref{equation: upper bound update}). This completes the proof of \textbf{Theorem \ref{theorem: update}}.

\ifCLASSOPTIONcaptionsoff
  \newpage
\fi

\bibliographystyle{IEEEtran}
\bibliography{summary}

\end{document}